\pdfoutput=1
\documentclass[fleqn,usenatbib]{mnras}
\usepackage{newtxtext,newtxmath}
\usepackage[T1]{fontenc}
\usepackage{ae,aecompl}
\usepackage{subfigure}

\usepackage{graphicx}	
\usepackage{amsmath}	

\newcommand{\cm}{$\rm cm$}

\title[Super-Eddington Emission from NSs]{Super-Eddington Emission from Accreting, Highly Magnetised Neutron Stars with a Multipolar Magnetic Field}

\author[N. Brice et al.]{
Nabil Brice,$^{1}$\thanks{E-mail: nabil.brice.17@ucl.ac.uk}
Silvia Zane,$^{1}$
Roberto Turolla,$^{2, 1}$
Kinwah Wu$^{1}$
\\
$^{1}$Mullard Space
Science Laboratory, University College London, Holmbury St. Mary,
Dorking, Surrey, RH5 6NT, UK \\
$^{2}$Department of Physics and Astronomy, University of Padova, via Marzolo 8, 35131Padova, Italy}

\date{Accepted 2021 March 26. Received 2021 March 10; in original form 2020 July 1}

\pubyear{2021}

\begin{document}
\label{firstpage}
\pagerange{\pageref{firstpage}--\pageref{lastpage}}
\maketitle

\begin{abstract}
Pulsating ultra-luminous X-ray sources (PULXs) are characterised by an extremely large luminosity 
($ > 10^{40} \text{erg s}^{-1}$).
While there is a general consensus that they host an accreting, magnetized neutron star (NS), 
the problem of how to produce luminosities $> 100$ times the Eddington limit, $L_E$, 
of a solar mass object is still debated. 
A promising explanation relies on the reduction of the opacities in the presence of a strong magnetic field,
which allows for the local flux to be much larger than the Eddington flux.
However, avoiding the onset of the propeller effect may be a serious problem.
Here, we reconsider the problem of column accretion onto a highly magnetized NS,
extending previously published calculations
by relaxing the assumption of a pure dipolar field
and allowing for more complex magnetic field topologies. 
We find that the maximum luminosity is determined primarily by the magnetic field strength
near the NS surface. 
We also investigate other factors determining the accretion column geometry
and the emergent luminosity,
such as the assumptions on the parameters governing the accretion flow
at the disk-magnetosphere boundary.
We conclude that a strongly magnetized NS with a dipole component of
$\sim 10^{13} \text{G}$, octupole component of $\sim10^{14} \text{G}$
and spin period $\sim1 \text{s}$ can produce a luminosity of $\sim 10^{41} \text{erg s}^{-1}$
while avoiding the propeller regime.
We apply our model to two PULXs, NGC 5907 ULX-1 and NGC 7793 P13,
and discuss how their luminosity and spin period rate
can be explained in terms of different configurations,
either with or without multipolar magnetic components.
\end{abstract}

\begin{keywords}
stars: neutron -- X-rays: binaries -- accretion, accretion discs
\end{keywords}



\section{Introduction}

Ultra-luminous X-ray sources (ULXs) are X-ray bright compact objects inside or near the optical extent of a galaxy but off-nucleus. Their observed X-ray luminosity exceeds the Eddington limit for a stellar mass, $M \sim 1 - 10 M_{\odot}$, compact object ($L > 10^{39} \text{erg s}^{-1}$). The nature of the compact object and the exact mechanism which powers the observed luminosity remains debated to this day (see \citealt{Kaaret2017} for the most recent review). 

For a long time, ULXs were thought to be either intermediate mass black holes \citep{Colbert1999} accreting at sub-Eddington rates or stellar mass black holes accreting at super-Eddington rates \citep{King2001}. The 
discovery of a pulsating ULX (PULX) M82 X-2 in 2014 by \cite{Bachetti2014} revolutionised our understanding of these sources. For the first time, firm evidence was presented that a ULX could host an accreting, neutron star (NS). Since then, further discoveries of PULXs have been made \citep{Furst2016, Israel2017a, Israel2017b, Rodriguez2019, Sathyaprakash2019}. At present, there is no 
evidence that PULXs differ from non-pulsating ULX basing on their spectral properties alone, which hints at the possibility that there are may be many more NS-powered ULXs than previously thought (e.g. \citealt{King2016}). This 
sparked a new theoretical effort aimed at investigating and modelling 
the physics of ultra magnetized 
accreting NSs, which is crucial to our understanding of ULXs.

The very first investigations of whether a 
strongly magnetized NS may be 
capable of emitting above its Eddington limit was presented several decades ago 
by \cite{Basko1976}. In their model, it is assumed that
the accretion disk is truncated far from the NS,
due to the interaction between the disk and the star magnetic field at the magnetospheric radius.
The accreting matter is then channelled along magnetic field lines onto the polar caps.
However, this particular model was primarily concerned with how the maximum luminosity can be affected by the presence of a funnelled accretion column and a radiative shock occurring above the star surface.
The model neglected the effects induced by the strong magnetic field on the plasma opacities
and on the radiation field and consequently,
\cite{Basko1976} found that the maximum luminosity could only be increased by a factor of a few above the Eddington limit.
This alone is insufficient to explain the super-Eddington luminosity observed in PULXs, e.g. in M82 X-2.
 
\cite{Lyubarskii1988} expanded the previous model by calculating the structure of the slow sinking region below the shock in two dimensions. This provided the basis for the most recent model by \cite{Mushtukov2015b}, which also includes the opacity reduction effect due to the NS's strong magnetic field. \cite{Mushtukov2015b} concluded that a maximum luminosity of up to $10^{40} \text{erg s}^{-1}$ can be sustained by the accretion column, depending on the strength of the magnetic field. However, problems arose in trying to explain the observed luminosity and spin period of NGC 5907 ULX-1 due to the assumption of a pure dipole field. Namely, using the model of \cite{Mushtukov2015b}, the required dipole magnetic field strength of NGC 5907 ULX-1 would place the source in the propeller regime \citep{Israel2017b}. In such a regime, the propeller effect is thought to halt the accretion process due to 
the transfer of angular momentum at the magnetopsheric radius from the NS to the accretion disk (see \citealt{Illarionov1975}).

In order to overcome this issue, \cite{Israel2017b} proposed that the pure dipole field assumption should 
be relaxed and that higher order multipole moments of the magnetic field may be 
dominant close to the surface of the NS. Such a magnetic field configuration is reminiscent of the one suggested for magnetars \citep{Turolla2015} and is supported by observational data from the magnetar SGR 0418+5729 \citep{Tiengo2013}, 
the isolated NSs RX J0720.4-3125 and J1308.6+2127 \citep{Borghese2015, Borghese2017},
and the millisecond pulsar PSR J0030+0451 \citep{Bilous2019}. A
multipolar magnetic field configuration would avoid the problem of the propeller effect induced by a super-strong dipolar
component while also allowing for sufficient opacity reduction
and in turn the release of a substantial super-Eddington luminosity. 

In this paper, we construct a model of column 
accretion onto a NS that allows for a multipolar magnetic field. 
As a basis for our calculation, we use the model described by \cite{Mushtukov2015b}. 
The paper is laid out as follows: 
in \S \ref{sec:accretion column model}, 
we give the basic equations and assumptions used in the model calculations. 
In \S \ref{sec:magnetic field}  
we present results of our numerical computations and compare with the previous model of \cite{Mushtukov2015b}. 
In \S \ref{sec:mixed polarization}
we present models based on different degrees of X-mode polarization,
while in 
\S \ref{sec:Disk Model}
we discuss how our results are affected by different assumptions on disk parameters. 
The robustness of some of our model assumptions and their regime of validity is discussed in 
\S \ref{sec:regime of validity}. 
The results of an application to two astrophysical sources,
which were suggested to have multipolar magnetic fields, is given in \S \ref{sec:application}.
Finally, we discuss our results and present our conclusions 
in \S \ref{sec:Discussion} and \S \ref{sec:Conclusions}.

\section{Accretion Column Model}
\label{sec:accretion column model}

Through the whole paper, 
we assumed a neutron star mass and radius of $M= 1.4 M_{\odot}$ and $R = 10^6$cm, respectively.

As a basis for our calculation, 
we consider the accretion column model originally developed by \cite{Basko1976}, 
according to which the free falling plasma is efficiently decelerated
in a radiative shock above the surface of the neutron star. 
Below the shock, 
the plasma slowly sinks toward the surface
and liberates its gravitational potential energy in the form of X-ray radiation. 
This region is referred to as the sinking region.

\subsection{Basic Equations}
\label{sec:radiative hydrodynamics}

For completeness and to introduce our notation, 
we summarise the radiative hydrodynamical equations 
that describe the plasma flow in the region below the radiative shock,
using the same assumptions as in \cite{Mushtukov2015b}. 

Since the accretion column is localized at the
magnetic poles and the height of the sinking region is less than the star radius, 
we neglect the curvature of the magnetic field lines and adopt an orthonormal coordinate system $(x,h)$,
with the $h$-axis along the magnetic field lines.
We define $x = 0$ to be the centre of the accretion column
and $h = 0$ to be at the surface of the NS.

\begin{figure}
\includegraphics[width=\columnwidth]{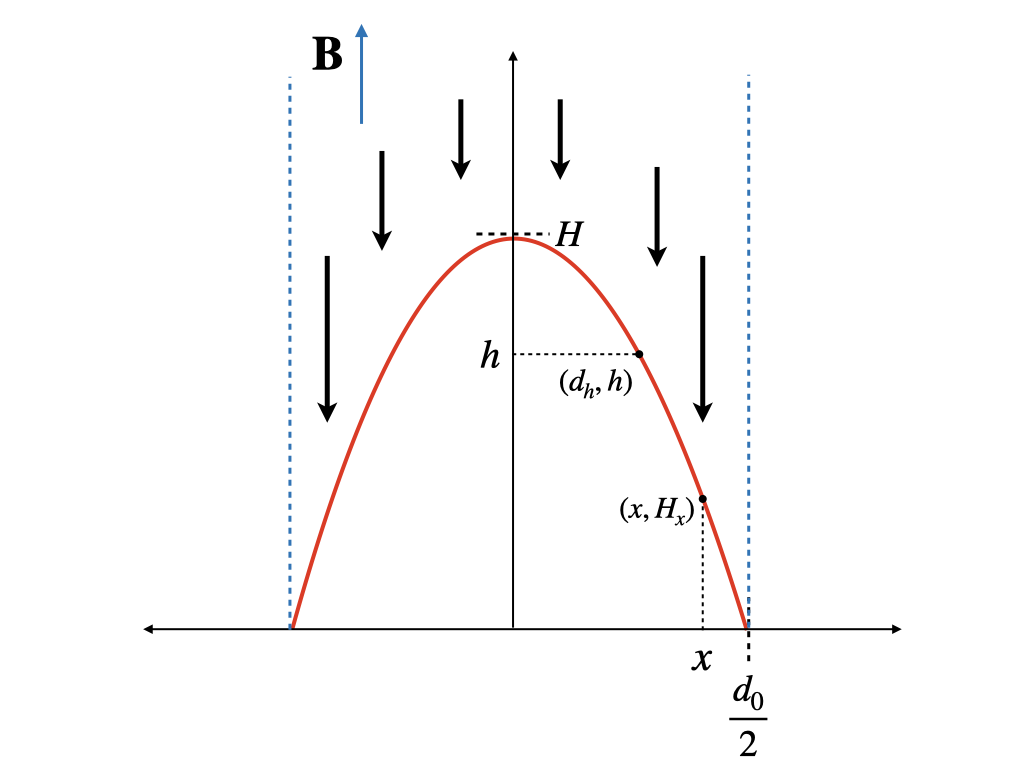}
\caption{
A vertical cross-section of the sinking region in the accretion column. 
The red line indicates the location of the shock in $(x,h)$ space. 
Below it, the plasma sediments towards the surface of the NS. 
The dashed blue lines represent magnetic field lines confining the accretion column. 
The maximum shock height is labelled $H$ and half the column base width is labelled as $d_0/2$. 
For a given $x$, the coordinates of the shock boundary are $(x, H_x)$. 
Alternatively, for a given $h$, the coordinates of the shock boundary are $\left(d_h/2,h\right)$.
}
\label{fig:columnillustration}
\end{figure}

 We indicate with $H$ the maximum height of the radiative shock,  and with $d_0$ the width of the accretion column base. $H_x$ is the height of the shock at width $x$ along the base and $d_h$ is the width of the sinking region at height $h$ above the surface. These geometrical quantities are illustrated in figure \ref{fig:columnillustration}, which shows a vertical cross-section of the sinking region. The footprint of the accretion column on the surface is an annulus with arc length $l_0$ and width $d_0$. The area of the accretion column base is given by $S_D = l_0 d_0$. The calculation of $l_0$ and $d_0$ is detailed in $\S$\ref{sec:Geometrical properties}.

Inside the sinking region, 
we consider a steady state flow, 
with velocity directed along the magnetic field lines only. 
A number of additional assumptions are made in order to simplify the equations further.
First, we assume that the radiation pressure, $P_{\text{rad}}$, dominates the gas pressure, $P_{\text{gas}}$. 
Following \cite{Mushtukov2015b}, 
we assume that the density and velocity profiles are independent of $x$, 
and coincide with the profiles at the center of the column (see $\S$\ref{sec:mass continuity}).
Finally, we assume the energy flux to be dominated by the radiative flux, which is a
the sum of two components,
one directed vertically along the field lines, $F_{\parallel}$, 
and one perpendicular to them, $F_\perp$.

Accordingly, the equations of continuity, momentum, and energy can be written respectively as 
\begin{align}
\rho v = \frac{\dot{M}}{2 S_D}, \label{eq:mass continuity} \\
v \frac{\partial v}{\partial h} + \frac{1}{\rho} \frac{\partial P_{\text{rad}}}{\partial h} +
\frac{G M}{(R + h)^2} &= 0, \label{eq:momentum conservation} \\
\frac{\partial F_{\parallel}}{\partial h} + \frac{\partial F_{\perp}}{\partial x} &= 0,
\label{eq:energy conservation}
\end{align} 
where $\rho$ is the plasma density, 
$v$ is the plasma velocity, 
$\dot{M}$ is the accretion rate, 
$M$ and $R$ are the mass and the radius of the neutron star, respectively. 
The energy fluxes $F_{\parallel}$ and $F_{\perp}$ can be expressed as 
\begin{align}
F_{\parallel} &= -\frac{c}{\kappa_{\parallel} \rho} \frac{\partial P_{\text{rad}}}{\partial h} +
P_{\text{rad}}v + uv + \rho v \left(\frac{v^2}{2} - \frac{GM}{R + h}\right), 
\label{eq:parallel flux} \\
F_{\perp} &= -\frac{c}{\kappa_{\perp} \rho} \frac{\partial P_{\text{rad}}}{\partial x},
\label{eq:perp flux}
\end{align}
where
$u$ is the radiation energy density, 
and $\kappa_{\parallel}$ and $\kappa_{\perp}$ are the angle and energy averaged Rosseland mean opacities 
in the direction parallel and perpendicular to the magnetic field lines, respectively (see $\S$\ref{sec:scattering cross-section}). 

In the sinking region below the shock, 
the flow is decelerated to a velocity much less then the free fall velocity,
$\frac{\partial}{\partial h} \left(v^2/2\right) \ll - GM/(R+h)^2$, 
so hydrostatic equilibrium can be assumed in the vertical direction, 
and thus equation (\ref{eq:momentum conservation}) is simply 
\begin{align}
\frac{\partial P_{\text{rad}}}{\partial h} = - \rho \frac{G M}{(R + h)^2} \label{eq:simple momentum conservation}  \ . 
\end{align}
Moreover, since the energy flux is dominated by the radiative flux, 
equations (\ref{eq:parallel flux}) and (\ref{eq:perp flux}) can be written as
\begin{align}
\frac{\partial P_{\text{rad}}}{\partial h} &= - \rho \kappa_{\parallel} \frac{F_{\parallel}}{c},
\label{eq:simple parallel flux} \\
\frac{\partial P_{\text{rad}}}{\partial x} &= - \rho \kappa_{\perp} \frac{F_{\perp}}{c}.
\label{eq:simple perp flux}
\end{align}

Equations (\ref{eq:simple parallel flux}) and (\ref{eq:simple perp flux}) can be integrated 
to calculate the radiation pressure distribution within the sinking region, 
and hence the whole structure of the latter. 
First, 
the parallel flux, $F_{\parallel}$, can be obtained by coupling equation (\ref{eq:simple parallel flux}) 
with the momentum conservation equation (\ref{eq:simple momentum conservation}). 
This yields
\begin{align}
F_{\parallel}(x,h) = \frac{c}{\kappa_{\parallel}} \frac{GM}{(R+h)^2}, 
\label{eq:Eddington flux}
\end{align}
which is the local Eddington flux, $F_{\text{Edd}}(x,h)$.
Then, 
the perpendicular flux, $F_{\perp}$, can be obtained integrating the energy conservation equation (\ref{eq:energy conservation}) in $x$ 
and by assuming $\frac{\partial F_{\parallel}}{\partial h} \approx \text{constant}$ in $x$. 
This yields
\begin{align}
F_{\perp} = F_{\perp, \text{esc}}(h)\frac{2x}{d_h}, \label{eq:Perpendicular Flux}
\end{align}
where $F_{\perp, \text{esc}}(h)$ denotes the perpendicular flux escaping from the sinking region at height $h$,
and we used the boundary conditions
\begin{align}
F_{\perp}(x = 0, h) = 0, \ F_{\perp}(x = d_h/2, h) = F_{\perp, \text{esc}}(h).
\label{eq:boundary conditions for perpendicular flux}
\end{align}
Equations (\ref{eq:simple parallel flux}) and (\ref{eq:simple perp flux}) can now be integrated,
yielding
\begin{align}
P_{\text{rad}, \parallel}(x, h) = \int_h^{H_x} \rho \frac{G M}{(R + y)^2} \ dy + \frac{2}{3}\frac{F_{\text{Edd}}(H_x)}{c}, 
\label{eq:Pradpara}
\end{align}
and 
\begin{align}
P_{\text{rad}, \perp}(x, h) = \frac{F_{\perp, \text{esc}}(h)}{c}\left[\frac{2}{d_h}
\int_x^{d_h/2} \rho \kappa_{\perp} z \ dz + \frac{2}{3}\right], 
\label{eq:Pradperp}
\end{align}
where $P_{\text{rad}, \parallel}$ and $P_{\text{rad}, \perp}$ are the radiation pressure
obtained by integrating the PDE in $h$ and $x$ respectively. 
Here, we used the boundary conditions
\begin{align}
P_{\text{rad}, \parallel}(x, h=H_x) = \frac{2}{3}\frac{F_{\parallel}(x, H_x)}{c} 
= \frac{2}{3}\frac{F_{\text{Edd}}(H_x)}{c}, 
\label{eq:boundary conditions for parallel radiation pressure} \\
P_{\text{rad}, \perp}(x=d_h/2, h) = \frac{2}{3}\frac{F_{\perp, \text{esc}}(h)}{c}.
\label{eq:boundary conditions for perp radiation pressure}
\end{align}

\subsection{Density Profile}
\label{sec:mass continuity}
The expressions for the radiation pressure in the sinking region, 
(\ref{eq:Pradpara}) and (\ref{eq:Pradperp}), depend on the plasma density, $\rho$.
The mass continuity equation 
(\ref{eq:mass continuity}) can be used to obtain the density profile in the sinking region, 
once the velocity profile is known. 
In principle,
we should solve the radiative hydrodynamical equations including the velocity terms
to obtain a fully self-consistent velocity profile.
Instead, following \cite{Mushtukov2015b}, 
we approximate the velocity profile by a power-law $v \propto h^{\xi}$, 
taking a fiducial value of $\xi = 1$. 
In \S\ref{sec:results}, we will discuss the effects of varying
the value of $\xi$ on the accretion column properties.

As previously mentioned in \S\ref{sec:radiative hydrodynamics}, 
we consider for simplicity the case in which the velocity profile is unchanged along $x$,
so that $v(x,h) = v(h)$.
As one of our boundary conditions,
we assume that the plasma is in free fall with velocity $v_{\text{ff}}$ above the shock,
and is decelerated to $v_{\text{ff}}(H)/7$ below the shock (therefore losing all but $\sim 1/50$ of its kinetic energy),
where $v_{\text{ff}}(H)$ is the free-fall velocity at height $H$.
We refer to \cite{Becker1998} for a one dimensional treatment of an adiabatic flow 
around the shock point that motivates the aforementioned velocity jump. 
The second boundary condition is given by the velocity vanishing at the surface of the NS.
Hence, at the height $h$ above the NS surface, 
the velocity is given by
\begin{align}
v(h) &= \frac{v_{\text{ff}}}{7} \left(\frac{h}{H}\right)^{\xi} \nonumber \\
&= \frac{1}{7} \sqrt{\frac{2GM}{R+H}} \left(\frac{h}{H}\right)^{\xi}. 
\label{eq:velocity profile}
\end{align}

By the mass continuity equation (\ref{eq:mass continuity}), we obtain the density profile
\begin{align}
\rho(h) = \frac{L_{\text{acc}}}{2S_D} \left(\frac{GM}{R}\right)^{-\frac{3}{2}} \left(\frac{49}{2}\right)^{\frac{1}{2}}  \left( 1 + \frac{H}{R} \right)^{\frac{1}{2}} \left(\frac{H}{R}\right)^{\xi} \left(\frac{h}{R}\right)^{-\xi}, \label{eq:density profile}
\end{align}
where we introduced the accretion luminosity $L_{\text{acc}}= \dot{M} G M/R$. 
Note that, as already pointed out by \cite{Mushtukov2015b}, formally the density diverges at $h=0$. 
Thus, the model assumptions become inadequate very close to the surface of the NS.
In particular, 
at some point the gas pressure will start to dominate over the radiation pressure.
In order to avoid this,
and following again \cite{Mushtukov2015b},
we truncate the numerical calculation slightly above the surface,
where $P_{\text{rad}} \approx P_{\text{gas}}$.

\subsection{Geometry of the accretion column}
\label{sec:Geometrical properties}
\begin{figure}
    \centering
    \includegraphics[width=\columnwidth]{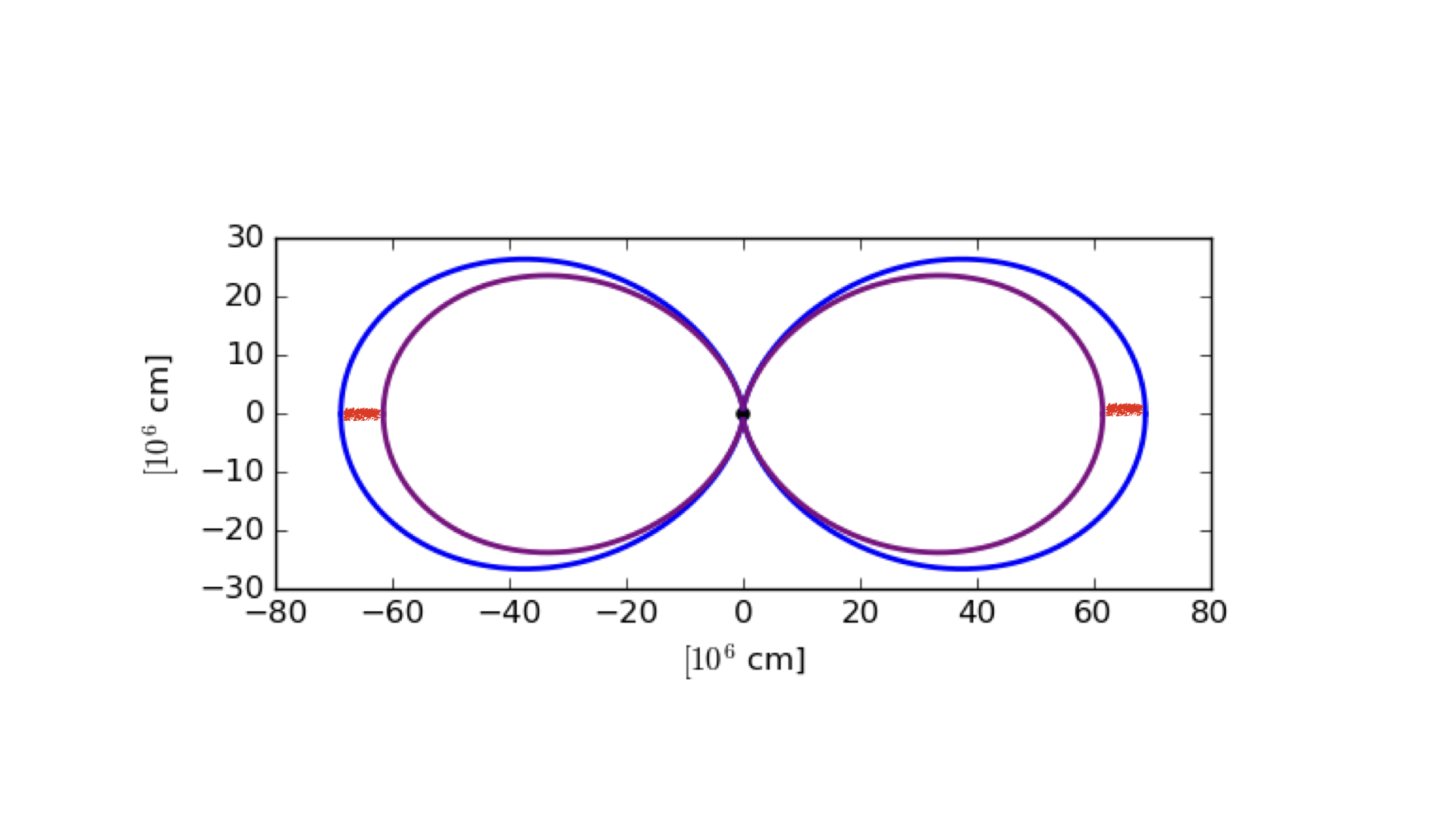}
    \caption{
    A 2D plot of the magnetic field lines, which have been computed numerically,
    in Cartesian coordinates centred on the NS.
    A multipolar magnetic field configuration,
    consisting of a dipole component
    and an octupole component $3$ times as strong as the dipole component,
    was used.
    Far from the NS ($h \gtrsim R$),
    there is no significant departure from the shape of a pure dipole magnetic field.
    The outer field lines (blue) are drawn in correspondence to the magnetospheric radius,
    while the inner ones (purple) to the inner radius of the boundary region.
    The two red segments show the part of the disk which enters the magnetosphere.} 
    \label{fig:example_magfield-png}
\end{figure}

The expressions for the radiation pressure in the sinking region, 
(\ref{eq:Pradpara}) and (\ref{eq:Pradperp}), 
explicitly contain terms related to the geometry, $H_x$ and $d_h$.
In addition, the density profile equation (\ref{eq:density profile}) 
contains the sinking region area $S_D$.
These geometrical quantities depend on the accretion column base geometry, $S_D(h=0) = l_0 d_0$, which in turn is determined by the specifics of the disk-magnetosphere interaction.
The simplest and most commonly used model is the one proposed by \cite{Ghosh1978},
according to which the disc is not sharply truncated at the magnetospheric radius.
The result is a boundary region with a finite width, although much smaller than the magnetospheric radius
(see figure \ref{fig:example_magfield-png}).
The two crucial quantities that appear in this simple disk-magnetosphere interaction model are 
the magnetospheric radius, $R_m$,
and the penetration depth, $P_m$, which determines the boundary region width.

The magnetospheric radius is given by
\begin{align}
R_m \approx 7\times10^7 \Lambda \ M^{1/7} R_6^{10/7} B_{\text{d}, 12}^{4/7} L_{39}^{-2 / 7} \text{cm}, \label{eq:magnetospheric radius}
\end{align}
where $\Lambda$ is a dimensionless parameter depending on the accretion mode, 
$R_6$ is the radius of the NS in units of $10^6\text{cm}$, 
$B_{\text{d}, 12}$ is the surface strength of the dipole component of the magnetic field in units of $10^{12}\text{G}$, 
and $L_{39}$ is the accretion luminosity in units of $10^{39}\text{erg s}^{-1}$. 
In this work we use the canonical value of  $\Lambda = 0.5$ \citep{Ghosh1978} for disk-fed accretion, and 
we investigate the effects of using different values of $\Lambda$ 
on the luminosity in \S\ref{sec:Disk Model}.
Besides, for the magnetic field configurations considered in this work,
the effects of including  higher order multipole components on the magnetospheric radius can be neglected,
as they decay much faster than the dipole component with increasing radius,
and typically $R_m \sim 100 R$.

The penetration depth, $P_m$,
is expected to be of the order of the disk height for a geometrically thin disk \citep{Ghosh1978}.
However, as it will be discussed later on, in certain models considered in this work,
the thin disk approximation is not valid,
and the parameters are more consistent with a geometrically thick disk.
In these cases, by taking a prescription that relates $P_m$ to the disk height would result in a penetration depth
greater than the magnetospheric radius,
which is physically unreasonable.
To account for this, we introduce an upper limit to the boundary region width
with respect to the magnetospheric radius. We introduce the penetration parameter  $\zeta \equiv  P_m / R_m$, and we assume $\zeta  \leq \zeta_{\text{max}}$, 
where $\zeta_{\text{max}}$ is a maximum penetration parameter. In the calculations,  
we use the prescription that the penetration depth is equal to the disk height,
following \cite{Mushtukov2015b},
and separate from this assumption only when $\zeta$ would be larger than the preset maximum
(see \S \ref{sec:Disk Model} for more details).
For our models, we use a fiducial value of $\zeta_{\text{max}} = 0.2$
(see \citealt{Li1999} for a study of the boundary region width).
However, in order to account for the uncertainty in the specifics of the disk-magnetosphere interaction, we also made some calculations 
by leaving $\zeta_{\text{max}}$ as a free parameter,
and in \S \ref{sec:Disk Model} we discuss the sensitivity of the accretion column base geometry
to changes in $\zeta_{\text{max}}$.

Once the penetration depth is set, the shape of the magnetic field lines constrains the accretion flow, and 
fully determines the accretion column base width, $d_0$, and length, $l_0$. Since, for general magnetic field topologies, the equation of the magnetic field lines cannot be expressed analytically, we  compute their shape numerically.
This is done by finding the vector expressions for the multipolar magnetic field
in polar coordinates
and proceeding to integrate in the direction parallel to the magnetic field lines
until reaching the surface of the NS.
A graphic of the result of one such calculation,
involving a multipolar magnetic field,
is shown in figure \ref{fig:example_magfield-png}.

The footprint of the magnetic field lines  
which also pass through the disk-magnetospheric boundary region forms an annulus centred on the magnetic axis with width $d_0$.
The accretion column base length, $l_0$, is then given by the mean of the inner circumference
and the outer circumference.

In addition to the column base geometry, 
the sinking region geometry depends on the shape of the magnetic field lines close to the surface
($0 \leq h \lesssim R$).
This must be taken into account when calculating the sinking region width above the surface, $d_h$,
as well as the accretion column area above the surface, $S_D$.

In the case of a pure dipole magnetic field, 
it is
\begin{align}
    d = d_0 \left(1 + \frac{h}{R}\right)^{3/2}, \ l = l_0 \left(1 + \frac{h}{R}\right)^{3/2},
    \label{eq:accretion-column-dimensions-dipole}
\end{align}
where $d$ and $l$ are the accretion column width and length, respectively, 
at a height $h$ above the surface.
As stated above, in the case of a multipolar magnetic field,
a numerical computation is required.
However, we find that the accretion column dimensions can still be written in the form
\begin{align}
    d = d_0 \left( 1 + \frac{h}{R} \right)^{\alpha}, \ l = l_0 \left( 1 + \frac{h}{R} \right)^{\beta},
    \label{eq:accretion-column-dimensions-multipole}
\end{align}
with the area thus being given by
\begin{align}
    S_D = l_0 d_0 \left( 1 + \frac{h}{R} \right)^{\alpha + \beta}.
    \label{eq:accretion-column-area-multipole}
\end{align}

\begin{figure}
    \centering
    \includegraphics[width=\columnwidth]{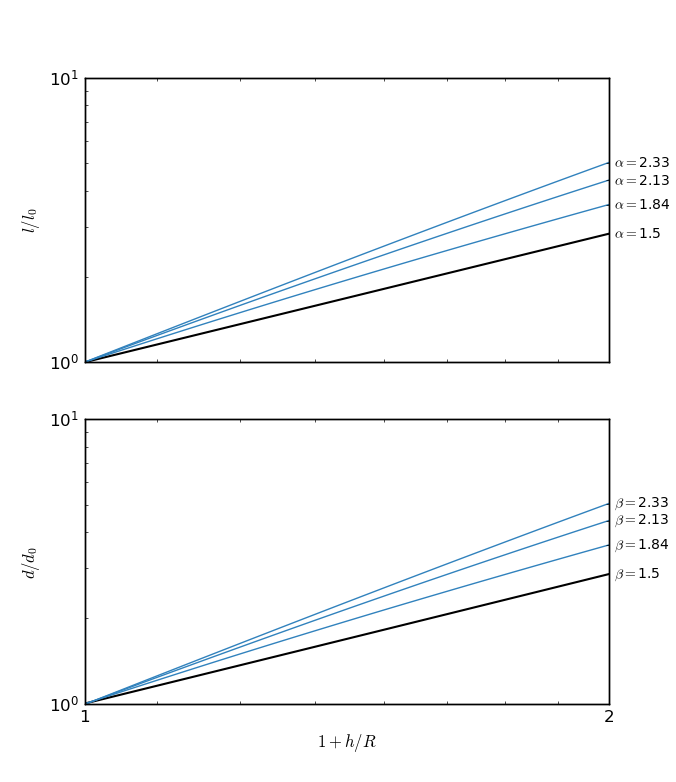}
    \caption{ The dimensions of the accretion 
    column against the height above the NS surface. 
    The horizontal axis shows the quantity $1 + h/R$, 
    to make the relationship outlined in equations
    (\ref{eq:accretion-column-dimensions-multipole}) manifest. 
    From the bottom to the top, the colored curves  
    are calculated in the case of a magnetic field with 
    a pure dipole component only, or with 
    $B_{\text{oct}} = B_{\text{dip}}$, 
    $B_{\text{oct}} = 3 B_{\text{dip}}$, 
    and $B_{\text{oct}} = 10 B_{\text{dip}}$, respectively.}
    \label{fig:landdvsh-png}
\end{figure}

In figure \ref{fig:landdvsh-png},
we show the variation of the accretion column dimensions,
for a selection of multipolar magnetic field configurations.
The values of $\alpha$ and $\beta$ in the multipolar magnetic field case
can differ considerably with respect to those of the pure dipolar case.

\subsection{Scattering opacity}
\label{sec:scattering cross-section}
The expressions for the radiation pressure in the sinking region,
(\ref{eq:Pradpara}) and (\ref{eq:Pradperp}),
contain the Rosseland mean opacity terms $\kappa_{\parallel}$ and $\kappa_{\perp}$,
which are the angle and frequency averaged opacities
in the directions parallel and perpendicular to the magnetic field lines, respectively.

In a strongly magnetised plasma, 
and assuming large Faraday depolarization, 
radiation propagates in two normal modes, 
the ordinary (O) and the extraordinary (X) mode,
with different polarization and opacity properties (see e.g. \citealt{Meszaros1992, Harding2006}). 
In this paper,    
we consider a pure scattering medium 
and calculate the electron scattering opacities in the (magnetic) Thomson limit, 
neglecting both the ion and vacuum contributions. 
The accreting plasma in the sinking region is assumed to be cold 
($k_{\text{B}}T \ll m c^2$ and $|E - E_{\text{cyc}}| \gg E \left(2 k_{\text{B}}T / m c^2 \right)^{1/2} | \cos (\theta) |$, 
where $m$ is the mass and $T$ is the temperature of the electrons). 
We use the expression for the frequency and angle dependent electron scattering opacity of the two modes 
as discussed in \cite{Kaminker1982} (see also \citealt{Zane2000}), 
and we assume a fully ionized solar mix plasma with mean molecular weight  $\mu_e = 1.17$.

The Rosseland mean opacity parallel to the magnetic field lines is given by
\begin{align}
\frac{1}{\kappa^i_{\parallel}} = \frac{\int_0^{\infty} \frac{\partial B_{E}(T)}{\partial T} dE \int_0^1 d\mu \ 3 \mu^2 \frac{1}{k^i(E, \mu)}}{\int_0^{\infty} \frac{\partial B_{E}(T)}{\partial T} dE }, \label{eq:parallel Ross mean}
\end{align}
and the Rosseland mean opacity perpendicular to the magnetic field lines is given by
\begin{align}
\frac{1}{\kappa^i_{\perp}} = \frac{\int_0^{\infty} \frac{\partial B_{E}(T)}{\partial T} dE \int_0^{\pi} d\varphi \int_0^1 d\mu_{\perp} \frac{3}{\pi} \mu_{\perp}^2 \frac{1}{k^i(E, \mu)}}{\int_0^{\infty} \frac{\partial B_{E}(T)}{\partial T} dE }.
\end{align}
Here $k^i(E,\mu)$ is the electron scattering opacity integrated over all possible outgoing photon directions 
(see the appendix of \citealt{Zane2000} for further details), 
the index $i$ denotes the polarization mode, 
where $i = 1$ is the X-mode and $i = 2$ is the O-mode, 
$B_E(T)$ is the Planck function, 
$E$ is the photon energy, 
$\mu$ is the cosine of the angle between the photon propagation direction and the magnetic field lines, 
and $\mu_{\perp}$ is the cosine of the angle between the photon propagation direction
and the direction perpendicular to the magnetic field lines. 
$\mu_{\perp}$ is related to $\mu$ by 
$\mu = \sqrt{1 - \mu_{\perp}^2} \cos \phi$,
where $\phi$ is the azimuthal angle relative to $\mu$. 

Following \cite{Mushtukov2015b}, 
we estimate the effective opacity for mixed polarization modes by 
\begin{align}
\frac{1}{\kappa} = \frac{f}{\kappa^1} + \frac{1 - f}{\kappa^2}, 
\label{eq:effective opacity}
\end{align}
where $f$ is the fraction of radiation in the X-mode. 
To make comparisons with the purely dipolar model presented by \cite{Mushtukov2015b},
we adopt the same approach of considering an accretion column with X-mode photons only ($f = 1$).
We discuss the variation of accretion column properties with a change in the X-mode fraction in 
\S \ref{sec:mixed polarization}.

Note that $\kappa_{\parallel}^i$ and $\kappa_{\perp}^i$ depend on the temperature, $T$, 
of the plasma. 
Since the sinking region is optically thick, 
we can approximate the plasma to be in thermal equilibrium with the radiation field. 
We relate the radiation pressure $P_{\text{rad}}$ to the temperature locally 
using the Eddington approximation 
and the blackbody approximation,
\begin{align}
P_{\text{rad}} \approx \frac{u}{3} \approx \frac{a T^4}{3}, 
\label{eq:LTE}
\end{align}
where $a$ is the radiation constant. 
Thus, the Rosseland mean opacities are calculated once the radiation pressure is known 
(see \S\ref{sec:Computation scheme} for an outline of the computation scheme).

\subsubsection{Scattering opacity for a hot plasma}
The cold plasma approximation is valid 
only when the thermal motions of the electrons are negligible compared with the phase velocity of the wave \citep{Harding2006}. 
While this is a sound assumption
for many of the photon energies and plasma temperatures 
encountered in the models studied here, 
there are a few cases for which the cold plasma approximation no longer holds. 
To estimate the effects of a hot plasma on the scattering opacity, 
we average the cold plasma scattering opacity with the thermal motions of the electrons
and introduce a line broadening of the cyclotron resonance.
This treatment has the advantage of providing an approximation
without resorting to a full computation of the magnetic Compton cross-section 
(which is beyond the purpose of this investigation).

In our treatment of the thermal motions of the electrons,
we consider the electron velocity distribution to be a one dimensional relativistic Maxwellian,
given by
\begin{align}
    f(p;T) \propto \exp\left[{-\frac{mc^2}{k_{\text{B}}T} \left(1 + \frac{p^2}{m^2 c^2}\right)^{\frac{1}{2}}}\right], 
    \label{eq:relativistic maxwellian}
\end{align}
where $p$ is the electron momentum along the magnetic field lines, 
and the proportionality constant is given by imposing the normalization condition, 
\begin{align}
    \int^{\infty}_{-\infty} f(p;T) \ dp = 1.
    \label{eq:normalization-condition}
\end{align} 
The averaged scattering cross-section is given by
\begin{align}
    \sigma(E, \mu, T) = \int_{-\infty}^{\infty} f(p;T) \ (1 - \mu \beta) \ \sigma_{\text{rf}}(E_{\text{rf}},\mu_{\text{rf}}) \ dp, 
    \label{eq:averaged scattering cross-section}
\end{align}
where $E_{\text{rf}}$ is the energy of the photon in the rest frame of the electron 
and $\mu_{\text{rf}}$ is the incident angle that the photon makes with the magnetic field lines in the rest frame of the electron. 
$\beta = v/c$ is the dimensionless electron velocity. 

For line broadening,
we increase the resonance width by adding
\begin{align}
    \Gamma = \left( 2 \frac{k_{\text{B}}T}{m c^2} \mu^2 \right)^{1/2} \label{eq:damping}
\end{align}
to the resonance damping term.
The full calculation is beyond the scope of this paper (see \citealt{Meszaros1992}).

Following these estimations, 
the hot plasma scattering opacity is frequency- and angle-averaged to obtain the Rosseland mean opacities, as described in \S \ref{sec:scattering cross-section}.

\subsection{Model Estimates}
\label{sec:Analytical Model}
An estimate of the radiation pressure, shape,
and luminosity of an accretion column can be made for a sinking region
with constant density profile $\rho(h) = \rho$,
and constant parallel and perpendicular opacity $\kappa_{\parallel}, \ \kappa_{\perp}$.
In this case, $P_{\text{rad}, \parallel}$ and $P_{\text{rad}, \perp}$ 
can be expressed analytically as
\begin{align}
P_{\text{rad}, \parallel}(x, h) = \rho \frac{GM}{R} 
\left[ \frac{H_x/R - h/R}{(1 + h/R)(1+H_x/R)}
+ \frac{2}{3} \frac{1}{\rho R \kappa_{\parallel}}\frac{1}{(1+H_x/R)^2}
\right], 
\label{eq:analytic Pradpara}
\end{align}
and  
\begin{align}
P_{\text{rad},\perp}(x,h) = \frac{F_{\perp, \text{esc}}(h)}{c} 
\left[ \rho \kappa_{\perp} d_h/4 \left( 1 - (2x/d_h)^2 \right) + 2/3
\right],
\label{eq:analytic Pradperp}
\end{align}
where only the functions $H_x$ and $d_h$ are left to be determined.
Note that $\rho \kappa_{\parallel} R$ is approximately the vertical optical depth 
of the sinking region,
while $\rho \kappa_{\perp} d_h/2$ is the horizontal optical depth of the sinking region at $h$.

The expression for the normalised escaping flux can be obtained
by equating the radiation pressure values obtained through equations
(\ref{eq:analytic Pradpara}) and (\ref{eq:analytic Pradperp}) computed at $x = 0$.
For $h = 0$, this yields
\begin{align}
\frac{F_{\perp, \text{esc}}(h = 0)}{c} = \rho \frac{GM}{R} \left[ \frac{ \frac{H/R}{(1+H/R)}
+ \frac{2}{3}\frac{1}{\rho \kappa_{\parallel} R}\frac{1}{(1+H/R)^2}}{\rho \kappa_{\perp} d_h/4 
+ \frac{2}{3}}
\right].
\label{eq:analytic Fperp}
\end{align}
Following \cite{Mushtukov2015b},
we now use  $P_{\text{rad},\parallel}(x, h=0) = P_{\text{rad},\perp}(x, h=0)$,
and assume $H_x/R \ll 1$, to obtain the relation  
\begin{align}
H_x/R \propto -x^2. \label{eq:analytic shape}
\end{align}
Hence, in this simplified case the shape of the shock is approximately quadratic near the base of the column and becoming less so near the top. 

The luminosity of the column is obtained by integrating the escaping flux over the surface of the column. Doing so yields
\begin{align}
    L = 4 l_0 \int_0^H \left( 1 + \frac{h}{R}\right)^{\beta} F_{\perp, \text{esc}}(h) \ dh,
    \label{eq:total-luminosity}
\end{align}
which is not integrable analytically due to the dependence of $d_h$ on $h$. 
However, we can obtain a lower bound for the luminosity by setting $d_h \approx d$,
i.e. approximating the horizontal optical depth of the accretion column with its maximum value. 
Since the column is optically thick,
i.e. $\rho \kappa_{\perp} d_h \gg 1$ and $\rho \kappa_{\parallel} R \gg 1$,
then the luminosity is approximately given by 
\begin{align}
    L \gtrsim \frac{4}{\pi} \left( \frac{l_0}{d_0} \right) 
    \left( \frac{\kappa_{\text{T}}}{\kappa_{\perp}} \right) f(H/R) L_{\text{Edd}},
    \label{eq:lum-lowerbound}
\end{align}
as also obtained by \cite{Mushtukov2015b},  where
\begin{align}
    f(H/R) = \frac{1}{1+H/R} \left[ (1+H/R)\log(1+H/R) - H/R\right].
    \label{eq:saturation-function}
\end{align}
However, note that the above equations are valid for any magnetic field configuration, provided $\alpha = \beta$, 
which for instance turns out to be the case when the magnetic field multipoles are aligned 
(as it is assumed in this work).
Equation (\ref{eq:lum-lowerbound}) gives an  approximate relation between the luminosity
and the accretion column base geometry.
As previously discussed by \cite{Mushtukov2015b}, we can also see that 
$f(H/R)$ grows only logarithmically for large $H/R$, and this sets a natural scale for the maximum luminosity at $H/R = 1$,
since the luminosity increases only marginally for higher $H$.

\subsection{Computational scheme}
\label{sec:Computation scheme}

The procedure for computing $P_{\text{rad}}$ from the radiation pressure equations,
(\ref{eq:Pradpara}) and (\ref{eq:Pradperp}), is non-linear
owing to the dependence of the opacity terms, $\kappa_{\parallel}$ and $\kappa_{\perp}$, 
on the plasma temperature, 
which itself depends on $P_{\text{rad}}$ through equation (\ref{eq:LTE}).
For this reason we use an iterative method,
and again we follow the scheme as given by \cite{Mushtukov2015b}, and we refer to this paper for all details.

We assume a magnetic topology of either purely dipolar or made up of a dipole plus an octupole component.  
The model parameters are the NS mass $M$, radius $R$, 
the accretion luminosity $L_{\text{acc}} = GM\dot{M}/R$, 
the velocity power-law index $\xi$, 
the polarization fraction $f$, 
the penetration parameter upper bound $\zeta_{\text{max}}$,
and the strength of each of the two magnetic field components at the NS surface. 
The total surface magnetic field strength at the poles, $B$, 
is the sum of the strength of each component at the poles, i.e. $B = B_{\text{dip}} + B_{\text{oct}}$. 

Subsequently, the magnetospheric radius $R_m$ is calculated according to equation (\ref{eq:magnetospheric radius}). 
The column base arc length $l_0$ and the column base width $d_0$ 
are calculated from $L_{\text{acc}}$ and $B$ 
by following the magnetic field lines from the magnetospheric radius to the surface of the NS 
(described in \S \ref{sec:Geometrical properties}).

For every set of parameters, the calculation is a double iterative process. 
First, we assume a trial value for the maximum shock height $H$, and we calculate iteratively the radiation pressure profile and the shape of the sinking region. 
This inner loop is repeated until the calculated  accretion column luminosity is within $1 \%$ from its value in the previous iteration.
At convergence, the luminosity corresponding to this trial value of $H$ is calculated by integrating the escaping flux over the surface of the accretion column:
\begin{align}
    L = 4 l_0 \int_0^H \left( 1 + \frac{h}{R}\right)^{\beta} F_{\perp, \text{esc}}(h) \ dh,
    \label{eq:luminosity}
\end{align}
A second iteration is then started, where  $H$ is then adjusted until the luminosity of the accretion column matches the accretion luminosity to within $1 \%$.

\section{Numerical Results}
\label{sec:results}

\subsection{Effects of the magnetic field strength and topology}
\label{sec:magnetic field}

In this section, we investigate how changes in the magnetic field strength and topology affect the accretion column properties, due to the changes they produce in the radiative opacities. We consider a magnetic configuration made of two components: a dipole and a higher order multipole. The dipole component dominates the behaviour of the field at large distances, i.e. at the magnetospheric radius, while the higher order multipole regulates the behaviour of the field near the NS surface. In the following calculations, we take the multipole moment to be the octupole. We chose the octupole moment over the quadrupole moment to better localize the effects of the surface magnetic field and avoid potential problems with null points of the magnetic field above the magnetic poles. In principle, other multipoles can be used. 

\begin{figure*}
    \centering
    \subfigure[]{\includegraphics[width=\columnwidth]{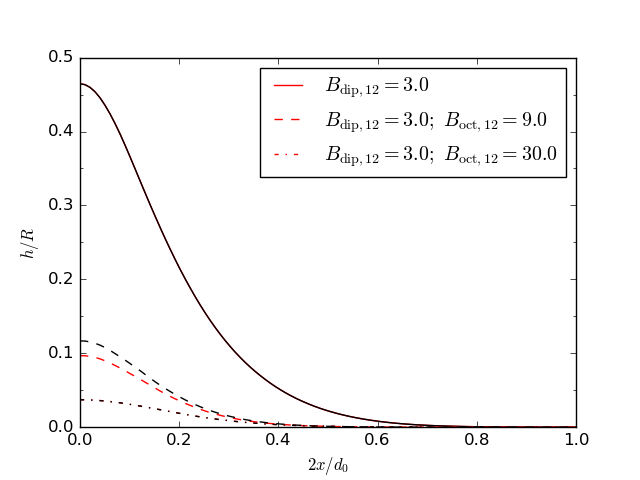}}
    \subfigure[]{\includegraphics[width=\columnwidth]{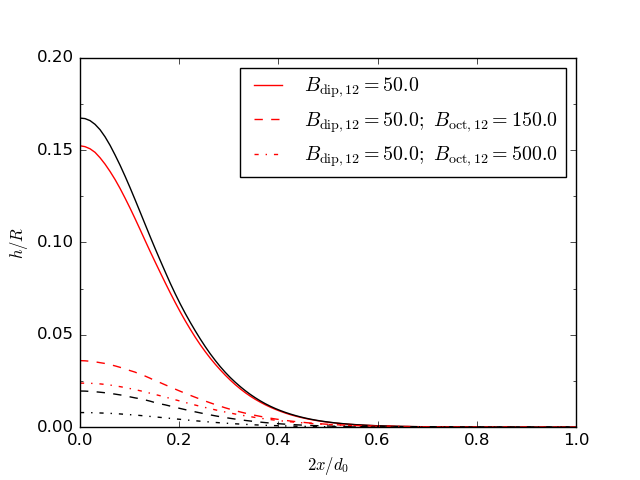}}
\caption{
Half vertical cross-section of the sinking region. In (a), we use $L_{39} = 1.0$, and the accretion column base variables were fixed at the values calculated for a pure dipole field with surface strength $3 \times 10^{12} \text{G}$, i.e.: $l_0 = 7.6 \times 10^5$ \cm\ and $d_0 = 1.4 \times 10^4$ \cm. In (b), we use $L_{39} = 10.0$, and $d_0 = 4.5\times10^4$\cm, and $l_0 = 4.7\times10^5$\cm, which correspond to a dipolar field of $5 \times 10^{13} \text{G}$. 
In both models the NS mass and radius are $M = 1.4M_{\odot}$, $R = 10^6$\cm, and the velocity power-law index is $\xi = 1$.
}
\label{fig:shapes}
\end{figure*}

\begin{figure*}
\centering
\subfigure[]{\includegraphics[width=\columnwidth]{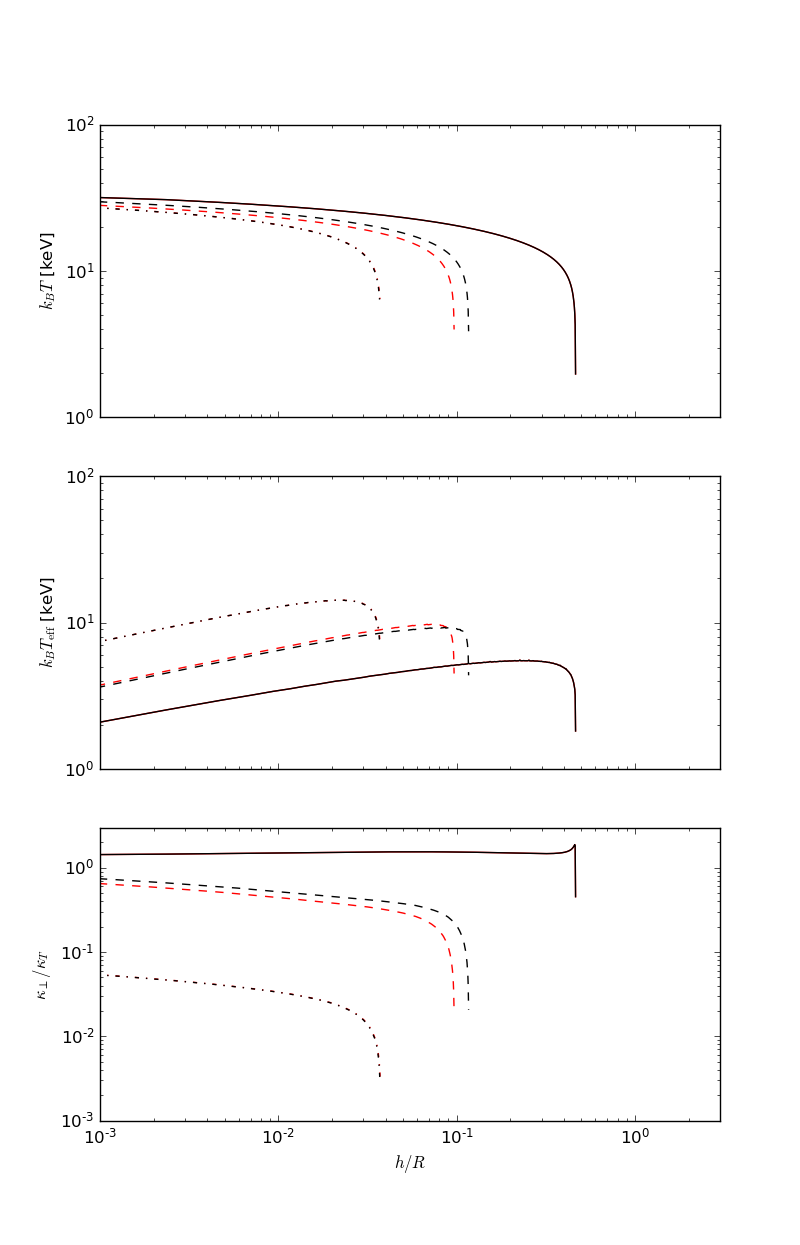}}
\subfigure[]{\includegraphics[width=\columnwidth]{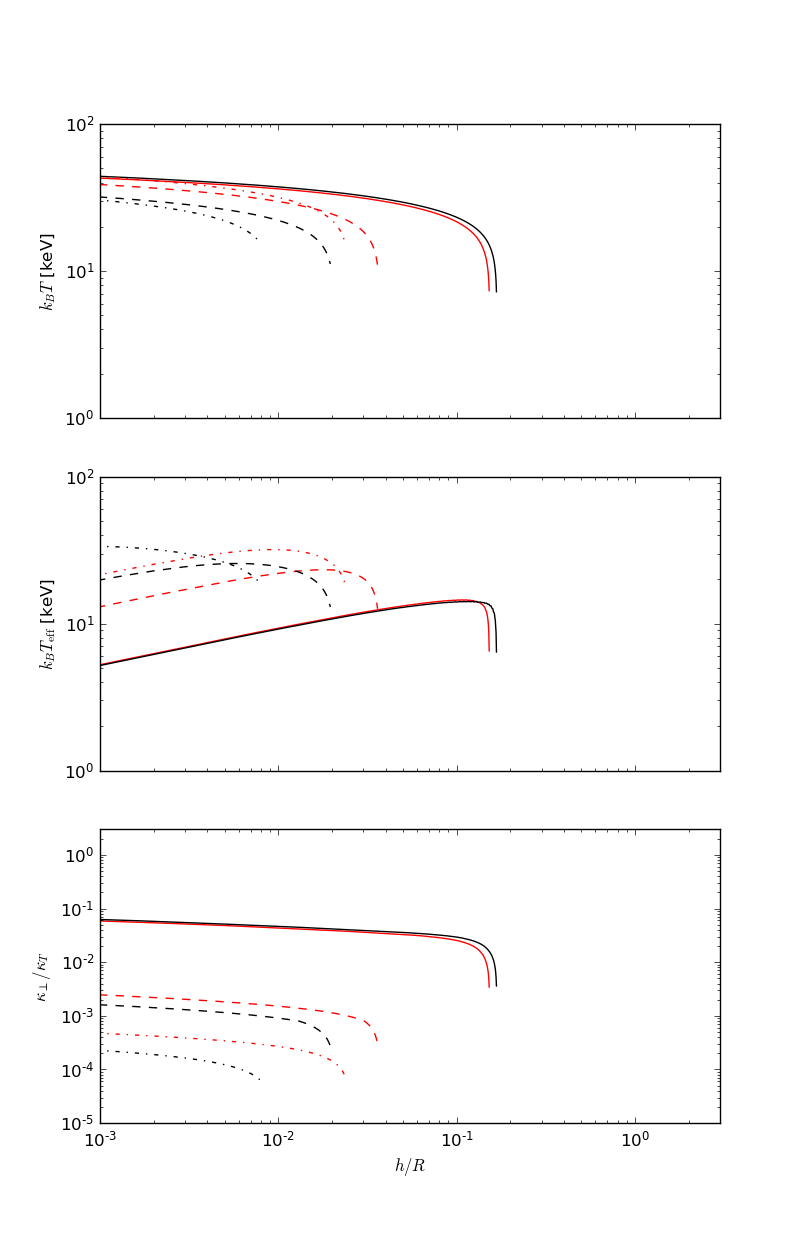}}
\caption{
Plots of properties of the accretion column sinking region, with (a) and (b) corresponding to the models shown in Fig.~\ref{fig:shapes}. From the top to the bottom, the different panels show: the internal ($x = 0$) temperature profile of the accretion column, the effective temperature of the emitted radiation, and the perpendicular mean opacity, $\kappa_{\perp}$ (see text for details).
}
\label{fig:properties1}
\end{figure*}

We start calculating a series of models by varying the strength of the octopolar component.
In order to separate the effects of the change in opacity from the changes in the column geometry (base and thickness) which are, in principle, also introduced by the magnetic field, we keep the accretion column base variables, $l_0$ and $d_0$, fixed to the values that they assume in the pure dipolar case.
For these models, we also neglect the curvature of the magnetic field lines.
Numerical results are presented in black in figures \ref{fig:shapes} and \ref{fig:properties1}
for a particular accretion luminosity and a velocity profile with power law index $\xi = 1$. We present two set of models, for two different values of the accretion luminosity and magnetic dipole strength ($L_{39} = 1.0$ and $B_{\text{dip}} = 3 \times 10^{12} \text{G}$, on the left, and $L_{39} = 10$ and 
$B_{\text{dip}} = 5 \times 10^{13} \text{G}$, on the right). In particular, in figure~\ref{fig:shapes} we show the vertical cross-section of the sinking region (only half of the region is shown, due to symmetry), to investigate the changes in the shock shape. Figures \ref{fig:properties1} shows the profiles of the central internal temperature, of the effective temperature, and of the perpendicular mean opacity in the accretion column. These properties allow an easy comparison with the model presented by \cite{Mushtukov2015b}, in which the field was assumed to be a simple dipole.

\begin{figure}
    \centering
    \subfigure[]{\includegraphics[width=\columnwidth]{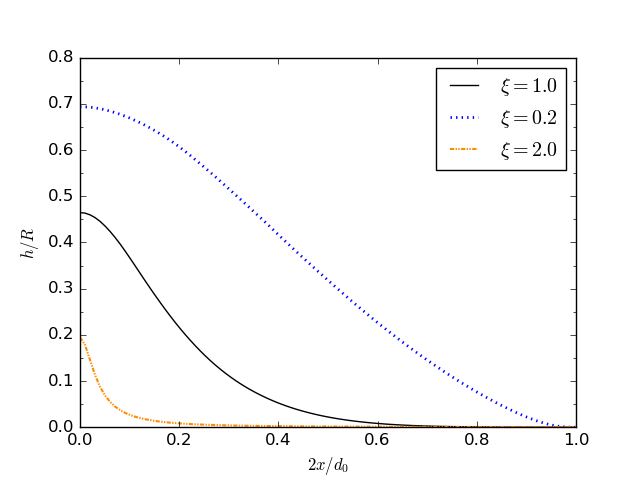}}
    \caption{ Half vertical cross-section of the sinking region for models with
    $L_{39} = 1.0$, $B_{\text{dip}} = 3\times10^{12} \text{G}$, and $B_{\text{oct}} = 0$. The different curves correspond to different values of the power law index: $\xi = 1$ (black solid line), $\xi = 0.2$ (blue dotted line), and $\xi = 2$ (orange dot-dashed line).}
    \label{fig:shapes_xichange}
\end{figure}

For a more self-consistent treatment of the effects of a multipolar magnetic field, $l_0$ and $d_0$ must also be allowed to vary. As discussed in $\S$ \ref{sec:Geometrical properties}, in a general case the base size and depth of the accretion column differs from those of a pure dipole magnetic field, so that $l_0$ and $d_0$ can both be reduced by a factor of several. In turn, this will affect the plasma density, the internal temperature, the escaping flux, and the maximum shock height. 
We therefore computed a set of models, by accounting for this effect and using the approach outlined in $\S$ \ref{sec:Geometrical properties} to calculate numerically $l_0$ and $d_0$ in correspondence of every assumed magnetic topology. 
For these models, the curvature of the magnetic field lines was taken into account.
Numerical results are shown in red in figures \ref{fig:shapes} and \ref{fig:properties1}, alongside the results for models with fixed column base geometry.

When using our numerical scheme we find that the maximum height of the shock, $H$, is slightly smaller than in the models presented by \cite{Mushtukov2015b}, which may be due to a difference in the calculation of the opacity (see \S\ref{sec:scattering cross-section}). However, the internal temperature profile, effective temperature profile, and opacity profile 
of our models (including those with an octupolar component) are qualitatively similar with those of \cite{Mushtukov2015b}. 
This indicates that the different field topology changes the quantitative details of the models but not the overall trend of the column properties.  

The first thing to note from figure \ref{fig:shapes} is that, as also pointed by \cite{Mushtukov2015b}, 
the shape of the shock is not quadratic in $x$, but instead the accretion column is quite narrow and the height of the shock above the surface drops practically to zero at a certain width $\Tilde{x} \lesssim d_0/2$ (we will refer to $\Tilde{x}$ as the ``sinking region width''). The sinking region width is the width at which the radiation pressure at the base becomes equal to the Eddington flux pressure, i.e. $P_{\text{rad}}(\Tilde{x}, h_0) = \frac{2}{3} F_{\text{Edd}}(h_0)/c$. For $x > \Tilde{x}$, the radiation pressure at the base is smaller than the Eddington flux pressure and no shock height can be supported.

The radiation pressure at $(x,h_0)$, i.e. at some distance along the column base, determines the height of the shock at $x$, $H_x$, by equation (\ref{eq:Pradpara}). Thus, the radiation pressure profile along the column base determines the shape of the shock. $P_{\text{rad}}(x, h_0)$ is determined by equation (\ref{eq:Pradperp}), in which we have assumed a linear function for the perpendicular flux, $F_{\perp}$, in our model. A different choice of function for $F_{\perp}$ will give a different shock shape. 

In figure \ref{fig:shapes_xichange} we present a series of models calculated varying the velocity index. 
As expected, a shallow velocity profile yields a shape more similar to that found for the analytical model, which predicts a quadratic column shape and is based on the simplified assumption of constant density. On the other hand, when a velocity index $\xi>1$ is used, the sinking region becomes narrower. This is a consequence of the greater deceleration of the particles in the lower layers of the accretion column. The upshot is that the radiation energy released by the particles is concentrated in the lower layers, which results in a lower shock height.

Figure \ref{fig:shapes} shows that the maximum shock height, $H$, decreases for an increasing surface magnetic field strength. By reversing the argument, for a fixed maximum shock height, a higher luminosity can be obtained by increasing the strength of the multipolar components. Increasing the magnetic field produces a larger opacity reduction in the X mode, therefore radiation escapes more readily from the sides of the sinking region. As a consequence, a smaller maximum shock height is sustained from the vertical radiation pressure. 

The internal central radiation temperature profile also shows an anti-correlation with the magnetic field strength (see the top plots of figure \ref{fig:properties1}). This is because models with a stronger magnetic field have a lower $H$, for reasons outlined previously. In models with a stronger magnetic field, particles start to be significantly decelerated by the shock at a point nearer to the NS surface. Hence by continuity, the density in the sinking region is lower, and in turn the internal radiation temperature is lower.  

The effective temperature, $T_{\text{eff}}$, obtained from the escaping flux using $F_{\perp, esc} = \sigma T_{\text{eff}}^4$, is also shown in figure \ref{fig:properties1}. As already noticed by \cite{Mushtukov2015b}, this quantity does not have a profile that simply reflects that of radiation temperature. Instead $T_{\text{eff}}$ increases with increasing height above the NS surface and then drops near the top of the accretion column. At the bottom of the accretion column, both the density, $\rho$, and the geometrical thickness of the sinking region, $d_h$, are large, which results in a large horizontal optical depth and a smaller escaping flux in that direction. Higher up, the accretion column becomes smaller in size, and the horizontal layers have a lower optical depth. The reduction in the optical depth is greater than that in the central temperature, and this is why the effective temperature generally increases with column height in the deeper regions. In fact, the peak of the effective temperature profile identifies the altitude at which the escaping flux is the greatest, and in turn this depends on the assumed density profile. For the accretion columns with velocity index $\xi = 1$, the effective temperature peak is close to the maximum shock height, where the optical depth is lower. For $\xi < 1$, the peak in $T_{\text{eff}}$ is at a lower altitude than when $\xi = 1$.

The perpendicular Rosseland mean opacity, $\Tilde{\kappa}_{\perp}$ (shown in figure \ref{fig:properties1}), is calculated at the central plane of the sinking region, $x = 0$. This quantity depends on both the total magnetic field strength in the accretion column and the temperature. In general, a higher magnetic field strength or lower temperature reduce the perpendicular Rosseland mean opacity. However, the perpendicular mean opacity is not a monotonic increasing function of temperature or magnetic field strength. In fact, it is largest when the photons in the sinking region have energies close to the electron cyclotron resonance energy.

Comparing the models with fixed column base geometry
(black lines in figures \ref{fig:shapes} and \ref{fig:properties1})
and models including the curvature calculation
(red lines in figures \ref{fig:shapes} and \ref{fig:properties1}),
it is immediate to note that there are no simple trends that explain the changes from the black curves to the red ones.
This is because the change in accretion column base geometry (which decreases when the strength of the multipolar component is increased)
results in a squeezing of the accretion column into a smaller area
while the change in the curvature of the magnetic field lines results in an increase of the accretion column area moving higher up the accretion column.
For each model, the overall outcome of these competing effects is different. 
However, it is worth noting that the change in the accretion column properties is modest and does not affect the qualitative behaviour discussed so far.

The effect of including the column geometry calculation is easier to explain in the models with multipolar magnetic fields shown
in figures \ref{fig:shapes}(b) and \ref{fig:properties1}(b),
which are low enough in height that the curvature does not make a substantial difference and the main effect is the reduction of the base size.
For these models (in red), the internal temperature at a given height in the sinking region is increased
compared with the fixed base geometry models (in black),
since the same amount of energy is produced in a smaller area,
and the perpendicular mean opacity rises following this increase in internal temperature.
In addition, the density of the sinking region is increased,
which makes it more difficult for radiation to escape from the sinking region,
hence decreasing the effective temperature.

On the other hand, the effect of the curvature alone can be understood by comparing the purely dipole models with $B_{\text{dip}}=5\times 10^{13}$~G  (figures \ref{fig:shapes}(b) and \ref{fig:properties1}(b)).
In this case, the accretion column base area is unchanged, since there are no multipolar magnetic fields, and the density is sufficiently high that the change in area due to the curvature is a consequential factor.
For this model (in red), the density of the accreting plasma is lower near the top compared with the fixed base geometry model (in black), and this subsequently decreases the perpendicular mean opacity,
allowing for more radiation to escape.
Hence, a lower maximum shock height is sustained. At lower fields (as for $B_{\text{dip}}=3\times 10^{12}$~G, figures \ref{fig:shapes}(a) and \ref{fig:properties1}(a)), the density inside the column is not high enough to make this  effect substantial.

\subsection{Mixed Polarization}
\label{sec:mixed polarization}

So far, we have considered only models with X-mode fraction $f = 1$. 
However, a more complete description of the radiation field in the sinking region
includes a mix of X-mode and O-mode photons
as well as scattering between polarization modes.
In particular, one may be worried that when a substantial fraction of O-mode photons are present,
the resulting decrease in the total opacity induced by the magnetic field
is not sufficiently large to allow for super-Eddington emission.
Solving this problem self-consistently requires a complete angle and frequency dependent solution
of the radiative transfer problem, which is not the purpose of this paper.
Instead, we investigate the issue by varying the value of $f$ in our calculation, i.e. we build up solutions by assuming that the radiation field consists of a given, fixed fraction of X and O-mode photons throughout the entire accretion column.

\begin{figure}
    \centering
    \includegraphics[width=\columnwidth]{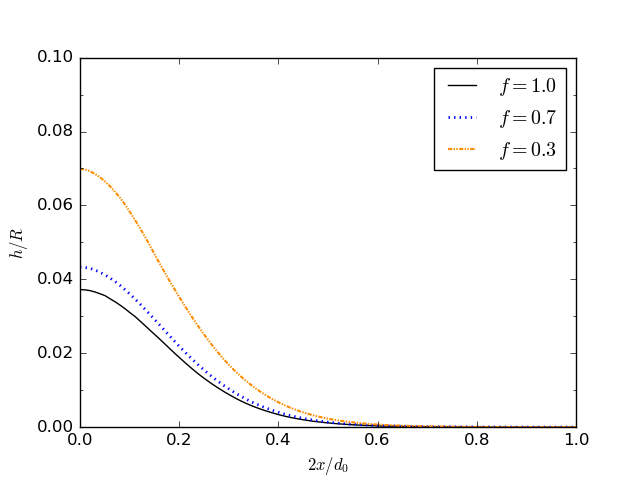}
    \caption{Half vertical cross-section of the sinking region.
    From top to bottom, the curves correspond to different values of the X-mode polarization
    fraction: $f = 0.3$ (orange dot-dashed line), 
    $f = 0.7$ (blue dotted line), 
    and $f = 1.0$ (black solid line).}
    \label{fig:shapes-f-png}
\end{figure}

\begin{figure}
    \centering
    \includegraphics[width=\columnwidth]{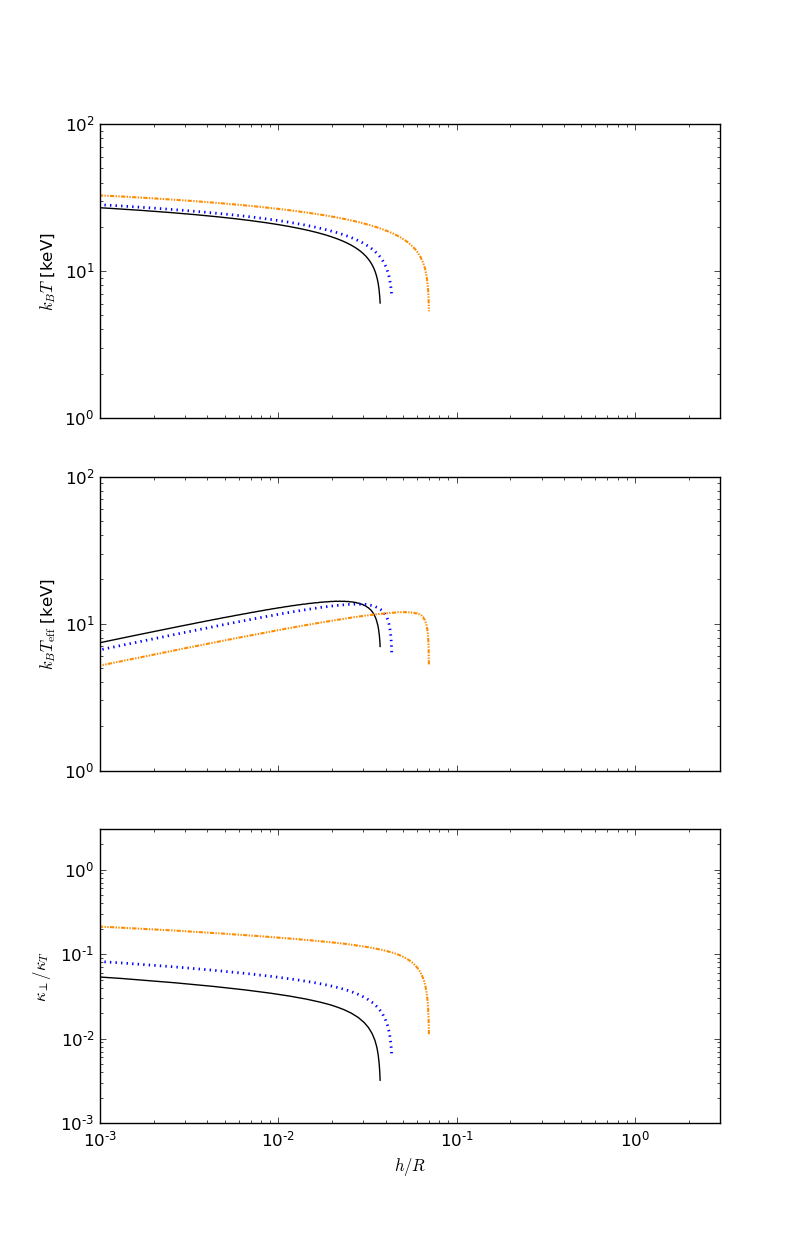}
    \caption{The accretion column properties for the same set of models as in Fig.~\ref{fig:shapes-f-png}.
    From top to bottom, the plots show
    the central internal temperature profile,
    the effective temperature profile of the emitting radiation,
    and the perpendicular mean opacity respectively.
    }
    \label{fig:properties-f-png}
\end{figure}

In general and as expected, we find that when a fixed fraction of O-mode photons is included, for an accretion column with a given magnetic field strength and accretion luminosity, the total opacity increases 
(as can be seen in the bottom plot of figure \ref{fig:properties-f-png}). However, even with a significant fraction of O-mode photons ($f = 0.3$),
the average perpendicular mean opacity is still well below
the Thomson scattering opacity in the cases considered here.
Solutions with a luminosity well above the Eddington limit are still possible, but since the opacity is larger, a higher shock height $H$ is sustained. This results in a higher internal temperature and lower effective temperature. 
A set of fixed X-mode fractions, $f = 1, \ 0.7,  \ 0.3$, was used to illustrate the effect of changing the polarization degree.  
In figure \ref{fig:shapes-f-png}, we show the vertical cross-section of the sinking region for an accretion column with magnetic field $B_{\text{dip}} = 3 \times 10^{12} \text{G}, \ B_{\text{oct}} = 3 \times 10^{13} \text{G}$ and  accretion luminosity $L = 10^{39} \text{erg s}^{-1}$. Figures \ref{fig:properties-f-png} show the internal temperature, effective temperature, and perpendicular effective opacity profiles respectively for an  accretion column 
computed by using the same parameters as in figure \ref{fig:shapes-f-png}. 

It is worth noticing that for low magnetic field strengths $(B < 10^{13} \text{G})$, the opacity decreases when a fraction of O-mode photons is included. This occurs when a large portion of the photons have energy close to or higher than the electron cyclotron resonance energy, $E_{cycl} \approx 11.6 \ B_{12} \ \text{keV}$. However, we are primarily interested in modelling sources with high magnetic field strengths $(B > 10^{13} \text{G})$, in which case most photons in the accretion column will have energies below $E_{cycl}$.

\subsection{Disk-magnetosphere interaction}
\label{sec:Disk Model}
In our modelling, calculation of the magnetospheric radius from equation (\ref{eq:magnetospheric radius})
requires an input parameter $\Lambda$,
while calculation of the penetration depth requires taking a particular prescription for $\zeta$
(see \S \ref{sec:Geometrical properties} for details).
However, both the exact value of $\Lambda$ and the expression for $\zeta$ are poorly known.
Thus, to test the robustness of the model results,
we studied the response of the accretion column base geometry variables, $l_0$ and $d_0$,
to changes in $\Lambda$ and $\zeta$ respectively.

For the models presented previously,
we used the canonical disk accretion value of $\Lambda = 0.5$ \citep{Ghosh1978}.
However, the exact value of $\Lambda$ depends on the extent to which the NS magnetic field threads
the accretion disk \citep{Wang1996},
and for instance \cite{Dall'Osso2016} suggest $\Lambda$ in the range $0.3 - 1$
as a conservative estimate of the possible values.
Repeating our calculations for $l_0$ with various $\Lambda$,
we find that $l_0$ changes less than an order of magnitude when $\Lambda$ is varied from 0.3 to 1.
Since $L \propto l_0$ and $\rho \propto S_D^{-1}$,
the overall accretion column properties are not very sensitive to changes in $\Lambda$.

With regards to $\zeta$,
we assumed a penetration depth proportional to the disk height at the magnetospheric radius,
as done by \cite{Mushtukov2015b}.
According to this prescription,
the penetration parameter is given by
\begin{align}
\zeta &= \frac{\kappa_{\text{T}}}{c} \frac{3}{8 \pi} \frac{\dot{M}}{R_m} 
\approx 0.2 L_{39}^{9/7} B_{\text{d}, 12}^{-4/7}.
\label{eq:ppMushtukov} 
\end{align}
However, since many of the models studied in this paper have a large accretion luminosity
(with $L_{39} \sim 10$)
and low dipole magnetic field strength (with $B_{\text{d},12} \sim 1$),
the penetration parameter in equation (\ref{eq:ppMushtukov})
can be close to or in excess of $\zeta = 1$.
Such values for $\zeta$ correspond to the disk penetrating through the entire magnetosphere
to the surface of the NS,
which is a physically unlikely scenario.
Thus, we introduced a maximum penetration parameter by hand (accordingly constraining $d_0$),
although the behaviour of $\zeta$ remains unchanged from equation (\ref{eq:ppMushtukov})
unless the disk becomes geometrically thick.

\begin{figure}
	\includegraphics[width=\columnwidth]{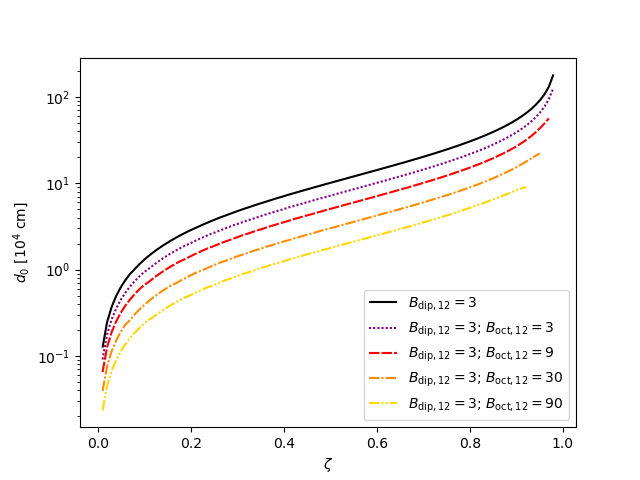}
    \caption{The accretion column base width, $d_0$, in units of $10^4$\cm\ for a given $\zeta$. We used $L_{39} = 1.0$, $M = 1.4 M_{\odot}$, $R = 10^6$\cm.}
    \label{fig:ZETAp0}
\end{figure}

Alternatively, a self-consistent approach would be to introduce a new prescription
for the penetration depth
based on some set of physical principles,
such as was done by \cite{Li1999}.
However, this would require extending previous disk-magnetosphere interaction models
to the case of a geometrically thick disk, which is not the purpose of this paper.
Hence, as a substitute to considering many different prescriptions,
we tested the response of $d_0$ to changes in $\zeta$ in general,
without assuming a particular disk-magnetosphere interaction model.
To do this,
we repeated the calculation for $d_0$
(see \S \ref{sec:Geometrical properties})
while varying $\zeta \in (0,1)$.
In each case, the accretion luminosity and magnetic field configuration are fixed. 
We considered $L = 10^{39} \text{erg s}^{-1}$
and several magnetic field configurations with $B_{\text{dip}} = 3 \times 10^{12} \text{G}$.
The results are reported in figure \ref{fig:ZETAp0}.

Our calculations show that across the domain of $\zeta$,
the change in $d_0$ can be an order of magnitude or more.
Since $F_{\perp, \text{esc}} \propto d_0^{-1}$ (see \S \ref{sec:Analytical Model}),
the luminosity of the accretion column is sensitive to changes in $\zeta$.
Hence, different prescriptions for the penetration depth
can lead to dissimilar results for
the accretion column properties, in particular the maximum luminosity.


\subsection{Maximum Luminosity}
\label{sec:maximum luminosity}
One of the central aims of this work is to investigate the maximum possible luminosity
from a highly magnetized, accreting NS, given some set of assumptions 
(such as the magnetic field configuration).
In order to calculate the maximum luminosity that can be sustained by the NS accretion column, 
we compute the maximum $L_{\text{acc}}$
for each set of model parameters by fixing the maximum shock height at $H = R$. 
In fact, at higher accretion column heights the luminosity only grows more slowly (see \S \ref{sec:Analytical Model}) and also the curvature of the magnetic field lines affects the vertical pressure balance equation (\ref{eq:momentum conservation}), making our approximation unsuitable. We repeated the calculation of the maximum luminosity for several magnetic field configurations, namely a pure dipole field, a field with $B_{\text{oct}} = 3 B_{\text{dip}}$, and a field with $B_{\text{oct}} = 10 B_{\text{dip}}$, 
while the other model parameters have been fixed at  $\xi = 1$, $f = 1$, $\zeta_{\text{max}} = 0.2$.

\begin{figure}
\includegraphics[width=\columnwidth]{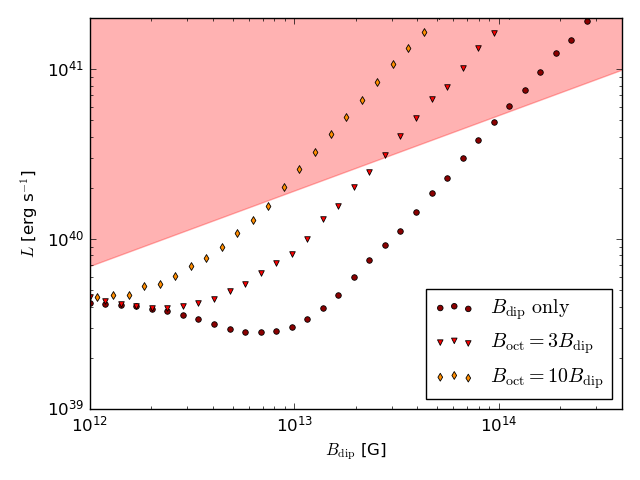}
\caption{
Maximum luminosity against surface dipole field strength.
The red shaded region indicates the region
for which $L$ exceeds the NS Eddington luminosity at the magnetospheric boundary,
i.e. when the accretion flow is super-Eddington.
The circles, triangles, and diamond points
show the computed maximum luminosity for 
a pure dipole,
a $B_{\text{oct}} = 3 B_{\text{dip}}$,
and a $B_{\text{oct}} = 10 B_{\text{dip}}$
surface magnetic field configuration respectively.
}
\label{fig:maxLplot}
\end{figure}

In agreement with the findings of \cite{Mushtukov2015b},
we find that, for total surface field strengths of $B < 10^{13} \text{G}$, the value of the maximum luminosity is mainly dictated by the accretion column geometry. This is because, 
at internal temperatures typical of the accretion column and for these low magnetic field strengths, 
most of the photons have $E > E_{\text{cycl}}$ 
and therefore are not subject to the reduction in opacity induced by the magnetic field. 
For higher total surface field strengths ($B > 10^{13} \text{G}$), 
the scattering opacity of the X-mode is instead significantly reduced
(by a factor of several to several orders of magnitude from $\kappa_{\text{T}}$)
such that this becomes the determining factor in constraining the maximum luminosity.
This can be seen by the change in slope of the maximum luminosity line in figure \ref{fig:maxLplot}.

The decreasing trend for the maximum luminosity for magnetic field strengths up to $10^{13} \text{G}$ can be explained by an increase in the temperature of the accretion column. Since the accretion column becomes thinner for higher magnetic field strengths (due to the choice of a maximum penetration paramater of $\zeta = 0.1$), the temperature increases, which also increases the overall scattering opacity.

As expected, when multipolar magnetic field configurations are accounted for,
we find that the maximum luminosity is increased when a stronger octupole component is present.
However this is simply due to the fact that the magnetic field increases in strength:
in fact, the maximum luminosity of a multipolar magnetic field corresponding to a total surface strength $B$ matches closely with the maximum luminosity for a purely dipolar magnetic field at the same $B$. 
Thus, the change in the column geometry due to the presence of higher order multipoles does not affect the maximum luminosity in a significant way.
Instead, as we will discuss in the next subsections, the maximum luminosity is more sensitive to the accretion column geometry
and hence to the prescription for the penetration depth into the magnetosphere.

\subsubsection{Maximum luminosity with mixed polarization}
\begin{figure}
\includegraphics[width=\columnwidth]{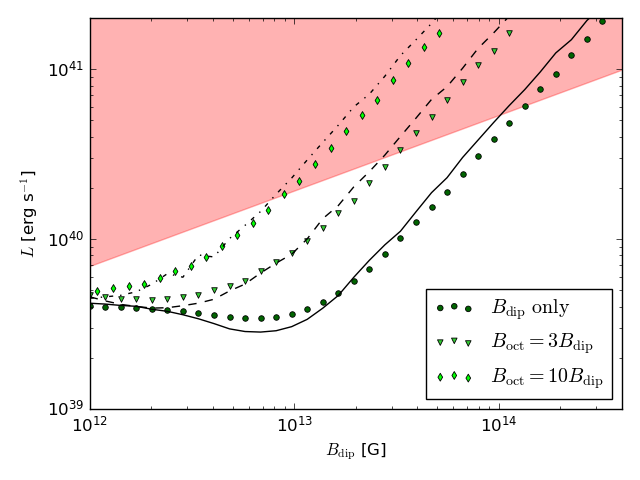}
\caption{ Same plot as in Fig.~\ref{fig:maxLplot} for models with different polarization fraction. 
The circles, triangles, and diamonds in green are computed using $f = 0.7$, and assuming  
a pure dipole,
a $B_{\text{oct}} = 3 B_{\text{dip}}$,
and a $B_{\text{oct}} = 10 B_{\text{dip}}$
surface magnetic field configuration, respectively.
The solid, dashed, and dot-dashed black lines shows the same configurations but computed with $f = 1$. 
}
\label{fig:maxLplot-f}
\end{figure}

Since the introduction of a mixed polarization radiation field changes the accretion column properties
(see \S \ref{sec:mixed polarization}), we also calculated the maximum luminosity for a fixed X-mode fraction of $f = 0.7$ (as might be expected in a more realistic case, for a scattering dominated model).  
Results are presented in figure \ref{fig:maxLplot-f}.
As the maximum shock height is typically increased,
the maximum luminosity is lower  than in the pure X-mode case. However, this trend is reversed for magnetic field strengths below $\sim 10^{13} \text{G}$.
This is due to a lowering of the average Rosseland mean opacity
when a fraction of O-mode photons is included, which 
occurs when a large portion of the photons have energy close to or higher than
the electron cyclotron resonance energy $E_{\text{cycl}} \sim 11.6 B_{12} \text{keV}$.

Otherwise, from figure \ref{fig:maxLplot-f}, 
the relation between the surface dipole field strength and the maximum luminosity
is seen to follow a shallower slope, as generally predicted.
However, the deviation is not too significant, less than a factor of $2$ in all cases we have examined.
This is due to the fact that most of the flux that supports the accretion column, in the diffusion approximation, is due to X-mode photons. Thus, the effective mixed mode opacity, calculated in Rosseland approximation, is dominated by the X-mode opacity.

\subsubsection{Comparisons with previous models}
\label{sec:maxL comparisons}

\begin{figure}
\includegraphics[width=\columnwidth]{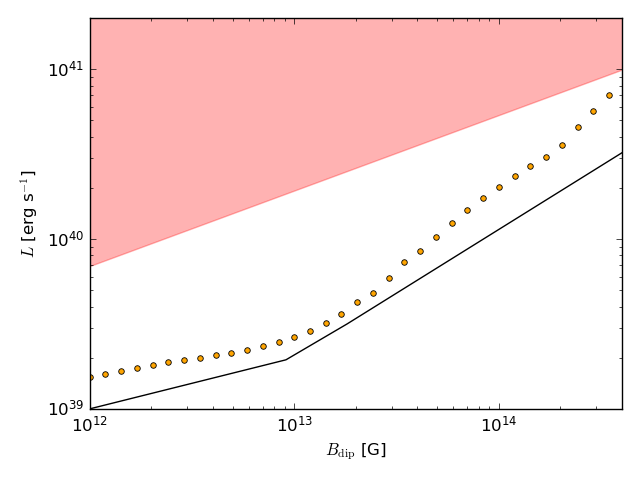}
\caption{ Same plot as in Fig.~\ref{fig:maxLplot} for models with a different disk model. 
The orange circles
show the computed maximum luminosity for 
a pure dipole surface magnetic field configuration, 
and with $\zeta_{\text{max}} \sim 1$.
The black line is the maximum luminosity
according to the relation given by \protect\cite{Mushtukov2015b}.
}
\label{fig:maxLcompplot}
\end{figure}

Compared with the model of \cite{Mushtukov2015b}, 
we have used a different method of calculating the scattering opacity
as well as a different disk-magnetosphere interaction model
(as described in \S\ref{sec:Disk Model}).
In figure \ref{fig:maxLcompplot},
we show the curve that represents the maximum luminosity obtained using our opacity files
and assuming the same disk model as \cite{Mushtukov2015b},
namely by using $\zeta_{\text{max}} \sim 1$.
In this case,
the maximum luminosity differs only by a factor of a few
with respect to the calculation presented by these authors,
indicating a good agreement between the two codes.

\subsection{Constraints on the parameter space}
\label{sec:regime of validity}

Before applying our accretion column model to observed PULXs, there are several considerations that we need to take into account. 
First, the region of the parameter space in which the model 
holds self-consistently is bound by the model assumptions,
and in primis
by the fact that we assumed a geometrically thin accretion disk at the magnetospheric boundary.
This means 
\begin{align}
H_m < R_m \, , \label{eq:Hm<Rm}
\end{align}
where $H_m$, the disk height at the magnetospheric radius, depends on the assumed disk model.
By using the standard thin accretion disk model of \cite{Shakura1973}, we find that,
for the strong accretion luminosities in which we are interested ($L > 10^{39} \text{erg s}^{-1}$),
the magnetospheric radius is always within the radiation pressure dominated zone
of the accretion disk.
In this zone, the disk height expression is given by
\begin{align}
H_m = \frac{\kappa_{\text{T}}}{c} \frac{3}{8\pi} \dot{M},
\label{eq:disk height}
\end{align}
which is independent of the radius. 

Using equations (\ref{eq:disk height}) and (\ref{eq:magnetospheric radius}),
the condition $H_m < R_m$ is equivalent to a lower bound on the dipole field strength,
above which our model is valid, which is given by 
\begin{align}
B_{\text{d}, 12} \gtrsim 0.034 \ L_{39}^{9/4} R_6^{-3/4} m^{-2} \Lambda^{-7/4}.
\label{eq:dipole lower bound}
\end{align}
For magnetic configurations with smaller dipole component,
the thickness of the disk becomes large at the magnetospheric boundary,
causing it to envelop the magnetosphere.
In this case, our estimates of the accretion column geometry
and our assumptions about the distribution of infalling plasma are no longer applicable.
A proper analysis of this scenario requires a new disk-magnetosphere interaction model,
which is beyond the purpose of this work. 

Second, since PULX are rotating NSs,
the strength of the dipole component must also be sufficiently small
so the propeller effect is avoided.
This means the magnetospheric radius must be smaller than the Keplerian corotation radius, so
\begin{align}
R_m < R_{co}, \label{eq:Rm<Rco}
\end{align}
The Keplerian corotation radius is given by
\begin{align}
R_{co} = \left( \frac{G M P^2}{4 \pi^2} \right)^{1/3}
\simeq 1.5 \times 10^8 m^{1/3} P^{2/3}\text{cm},
\label{eq:corotation radius}
\end{align}
where $P$ is the NS spin period.
Equation (\ref{eq:Rm<Rco}) can thus be written as
an upper bound for the dipole magnetic field strength,
\begin{align}
B_{\text{d}, 12} \lesssim 4.57 \ \Lambda^{-7/4} m^{-1/12} R_6^{-5/2} L_{39}^{1/2} P^{7/6}.
\label{eq:dipole upper bound}
\end{align}
For magnetic configurations with a larger dipole component,
the propeller effect prevents accretion onto the poles \citep{Illarionov1975}.

Third,
we will assume the spin period derivative to be dominated by the accretion torque.
A simple accretion torque model is used to estimate the minimum average accretion rate
that can give rise to the measured secular spin period derivative.
In this model,
we assume the angular momentum of the accreting matter is transferred to the NS
at the corotation radius,
which is the largest distance at which accretion can still occur.
This produces the largest torque on the NS for a given amount of accreted material
(and thus the lowest accretion luminosity).
Thus,
\begin{align}
- 2 \pi I \frac{\dot{P}}{P^2} < \dot{M} \sqrt{G M R_{co}},
\label{eq:accretion torque}
\end{align}
where $I \approx 10^{45} \text{g cm}^2$ is the moment of inertia of the NS, $P$ 
and $\dot{P}$ are the spin period and its derivative.
Equation (\ref{eq:accretion torque}) can then be rearranged
to give a lower bound on the accretion luminosity,
\begin{align}
L_{\text{acc}} = \dot{M} \frac{GM}{R} > 0.66 \ \dot{P}_{-10} P^{-7/3} 10^{39} \text{erg s}^{-1},
\label{eq:accretion luminosity lower bound from spin period derivative}
\end{align}
where $\dot{P}_{-10} = 10^{-10} \dot{P}$, and $P$ is in seconds. Values of the 
accretion luminosity lower then this limit would be  insufficient in explaining the observed $\dot P$, according to our simple model.

\subsection{Applications}
\label{sec:application}
Working with the parameter space restrictions derived in \S \ref{sec:regime of validity},
we can diagnose the necessity of higher order multipole magnetic field components
in observed astrophysical sources. 
We apply the model to two PULXs,
for which the face value application of the model by \cite{Mushtukov2015b}
has led to the suggestion of the presence of multipolar magnetic fields,
namely NGC 5907 ULX1 \citep{Israel2017b} and NGC 7793 P13 \citep{Israel2017a}. 
The luminosity of both of these sources show a variation by a factor of $\sim 8$,
which is large but still more likely to be due to a variation in the accretion rate
rather than a transition to the propeller effect.
Hence, we apply the dipole magnetic field strength upper bound condition
given by equation (\ref{eq:dipole upper bound})
for the entire luminosity range exhibited.

\subsubsection{NGC 5907 ULX-1}
\label{sec:NGC 5907 ULX-1}
NGC 5907 ULX-1 \citep{Israel2017b} is the brightest PULX found to date,
with a peak luminosity $L_{\text{peak}} = (2.3 \pm 0.3) \times 10^{41} \text{erg s}^{-1}$
and observed luminosity variation between $L = 2.6 \times 10^{40} \text{erg s}^{-1}$
and $L = 2.3 \times 10^{41} \text{erg s}^{-1}$.
Observations performed with {\it XMM-Newton} in 2003 and 2014,
have shown a decrease in the pulse period from $\sim 1.43 \text{s}$ to $\sim 1.137 \text{s}$,
which corresponds to a secular spin period derivative
$\dot{P} \approx - 8 \times 10^{-10} \text{s s}^{-1}$. 

\begin{figure}
\includegraphics[width=\columnwidth]{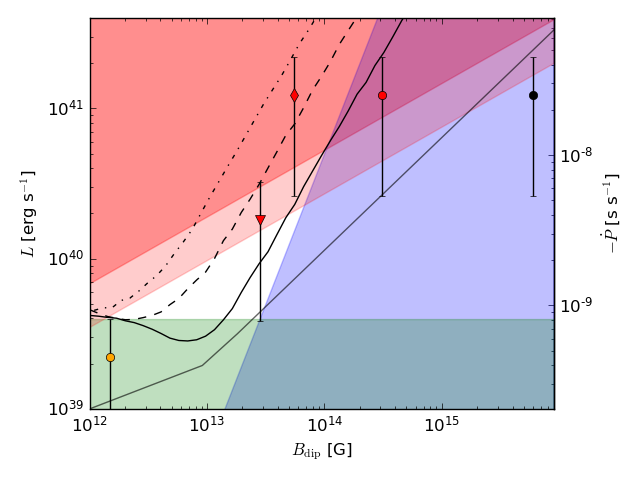}
\caption{
The parameter space plot of the magnetic field dipole component strength 
and the accretion luminosity 
for the source NGC 5907 ULX-1.
The light red and darker red shaded area indicate the region
for which $L$ exceeds the thick disk
and NS Eddington luminosity at the magnetospheric boundary, respectively.
The blue shaded area indicate region for which the source is in the propeller regime.
The green shaded area shows the region for which the accretion rate is too low
to provide sufficient secular spin period derivative $\dot{P} = - 8 \times 10^{-10} \text{s s}^{-1}$, 
calculated based on equation (\ref{eq:accretion luminosity lower bound from spin period derivative}).
The solid, dashed, and dash-dotted black lines show the maximum luminosity
in the case of a pure dipole field,
a field with $B_{\text{oct}} = 3 B_{\text{dip}}$,
and a field with $B_{\text{oct}} = 10 B_{\text{dip}}$ respectively. 
The solid gray line shows the maximum luminosity according to the relation given by \protect\cite{Mushtukov2015b}.
Several example configurations for the source are shown by the black, red, and orange dots.
In each case, the vertical lines represent the observed luminosity range
when the source was in a high luminosity state \protect\citep{Israel2017a}.
In particular, the black dot shows the configuration required to remain under the maximum luminosity
when using the relation given by \protect\cite{Mushtukov2015b}.
The red circle, diamond, and triangle show possible configurations
(pure dipole, $B_{\text{oct}} = 10 B_{\text{dip}}$, $B_{\text{oct}} = 3 B_{\text{dip}}$ respectively)
with beaming factors of $b = 1.0, \ 1.0, \ 0.1475$ respectively.
The orange circle shows a configuration with high radiation collimation ($b < 0.02$).
}
\label{fig:NGC5907ULX1}
\end{figure}

By discussing the source in the context of the \cite{Mushtukov2015b} model,
\cite{Israel2017b} suggested the need of multipolar magnetic components.
In order to test this argument, 
we plot again a figure analogous to Fig. 3 of \cite{Israel2017b}
but using our computed maximum luminosity.
The parameter space constraints in the $L$-$B_{\text{dip}}$ plane, 
and few example configurations for NGC 5907 ULX-1
are shown in figure \ref{fig:NGC5907ULX1}. 

In agreement with previous findings,
we find that in order to explain the whole range of observed luminosities
up to $L_{\text{peak}} \sim 10^{41} \text{erg s}^{-1}$ while assuming a pure dipole magnetic field,
a strong magnetic field strength is necessary,
i.e. $B > 10^{14} \text{G}$ (from our model) or even $B > 10^{15} \text{G}$
(from the model of \citealt{Mushtukov2015b}).
Specifically, our model suggests that a pure dipole field
of surface strength $B_{\text{dip}} \approx 3.1 \times 10^{14} \text{G}$
can give rise to a luminosity of $2.3 \times 10^{41} \text{erg s}^{-1}$.
Although the strength of this magnetic field is an order of magnitude lower than inferred
from the \cite{Mushtukov2015b} model,
the source would still be in the propeller regime (see figure \ref{fig:NGC5907ULX1}).
 
For the PULX to be emitting with $L_{\text{peak}}$ without entering the propeller regime, 
a multipolar magnetic field is required,
in particular a dipole component surface strength
of $B_{\text{dip}} \approx 5.5 \times 10^{13} \text{G}$
and octupole component surface strength of $B_{\text{oct}} \gtrsim 5.5 \times 10^{14} \text{G}$
(see figure \ref{fig:NGC5907ULX1}).
However, for all observed values of the luminosity,
this configuration falls within the thick disk regime,
i.e. the dipole magnetic field strength is lower than the bound given in equation (\ref{eq:dipole lower bound}). While it can not be excluded that this contradiction may be resolved once a different and less simplified disk model is
adopted, to answer this question would require a thorough treatment of disk accretion and disk-magnetosphere interaction,
which is beyond the scope of this paper. 

As it can be seen in figure \ref{fig:NGC5907ULX1},
both the super-Eddington disk accretion rate
and the propeller regime can be avoided
by introducing a moderate beaming factor of $b \lesssim 0.15$.
In this case, a model based on a multipolar magnetic field configuration
with dipole component surface strength of $B_{\text{dip}} \approx 2.8 \times 10^{13} \text{G}$
and a slightly larger octupole surface strength of $B_{\text{oct}} \gtrsim 8.4 \times 10^{13} \text{G}$
can explain the entire observed range of luminosities, up to $b L_{\text{peak}}$. 

Stronger beaming factors $b \lesssim 0.02$ allow for a reduced accretion luminosity
and therefore again make possible a pure dipole field configuration.
However, in this case, the average accretion luminosity falls below the threshold
required to give the secular spin period derivative $\dot{P} \approx 8 \times 10^{-10} \text{s s}^{-1}$,
i.e. the luminosity is lower than the bound given in equation
(\ref{eq:accretion luminosity lower bound from spin period derivative}). 
Hence, according to our model,
the most favourable configuration for NGC 5907 ULX-1 includes a moderate beaming factor
and crucially a multipolar magnetic field.

\subsubsection{NGC 7793 P13}
\label{sec:NGC 7793 P13}
\begin{figure}
	\includegraphics[width=\columnwidth]{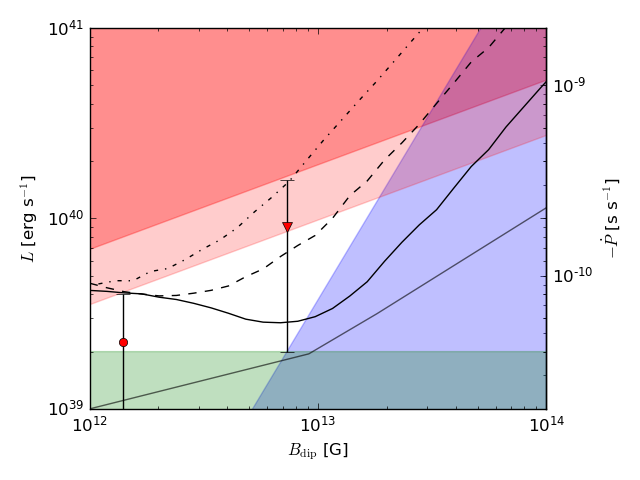}
    \caption{
    The parameter space plot of the magnetic field dipole component strength
    and the accretion luminosity for the source NGC 7793 P13.
    The shaded regions indicate the same regimes as in figure \ref{fig:NGC5907ULX1},
    except using $P = 0.42$ and $\dot{P} = -4.0 \times 10^{-11} \text{s s}^{-1}$.
    The different lines mark the maximum luminosity for the same magnetic field configurations as
    in figure \ref{fig:NGC5907ULX1}.
    The observed range of luminosities is shown in two configurations,
    with the red triangle and red circle assuming unbeamed emission and beamed model with $b = 0.25$ respectively. 
    }
    \label{fig:NGC7793P13}
\end{figure}

The PULX NGC 7793 P13 was observed 
to have a peak luminosity $L_{\text{peak}} = 1.6 \times 10^{40} \text{erg s}^{-1}$ 
and luminosity variation between 
$L \sim 2.0 \times 10^{39} \text{erg s}^{-1}$ and $L = 1.6 \times 10^{40} \text{erg s}^{-1}$. 
A spin period of $\sim 0.42 \text{s}$ was measured and
a secular spin period derivative $\dot{P} \sim -4.0 \times 10^{-11} \text{s s}^{-1}$ inferred
from two observations one year apart \citep{Israel2017a}.
Our constraints on the possible values of $L$ and $B_{\text{dip}}$ are shown,
together with some example configurations for this source,
in figure \ref{fig:NGC7793P13}. 

In this case, 
the whole range of observed luminosities,
up to the peak value of $1.6 \times 10^{40} \text{erg s}^{-1}$,
can be achieved by a configuration with
a multipolar magnetic field of dipole component surface strength $B_{\text{dip}} \approx 7.3 \times 10^{12} \text{G}$
and a much stronger octupole component with surface strength $B_{\text{oct}} > 7.3 \times 10^{13} \text{G}$
(see the red triangular point in figure \ref{fig:NGC7793P13}).
Under these conditions, the source is not in the propeller regime and no beaming 
is required to avoid the super-Eddington disk regime.
This particular configuration has the advantage
of being comfortably above the lower luminosity bound required to also explain the observed spin period derivative.
However, the largest observed flux levels are not compatible with the assumption of geometrically thin disk,
which again may demonstrate that disk model is over simplified.  

When a mild beaming factor $b \lesssim 0.25$ is introduced,
the effective luminosity $b L_{\text{peak}}$ becomes small enough
that it can be explained by a configuration with a pure dipole field
of surface strength $B_{\text{dip}} \sim 1.4 \times 10^{12} \text{G}$.
This is at variance with respect to the conclusions based on the calculation of \cite{Mushtukov2015b}.
The only downside is that the lowest observed luminosities
fall below the bound required to explain the observed spin period derivative.
On the other hand,
the secular spin period derivative may be the cumulative result of alternating accretion phases
and may have been accumulated during epochs of larger mass transfer.
Thus, this particular configuration is not severely disfavoured, according to the accretion torque model used here.

\section{Discussion}
\label{sec:Discussion}
Motivated by the recent discovery of pulsating ULXs 
\citep{Bachetti2014, Furst2016, Israel2017a, Israel2017b, Carpano2018, Rodriguez2019, Sathyaprakash2019}
and their proposed interpretation in terms of accreting magnetars (see \citealt{Tong2019}),
we have reconsidered the problem of columnated accretion onto a highly magnetized NS.
The main aim was to find model configurations capable of producing a high, super-Eddington luminosity
while not in the propeller regime
(for the values of the spin period typical of PULXs, $P \sim 1\text{s}$).

We worked in a scenario similar to the one recently discussed by \cite{Mushtukov2015b}
but we relaxed the assumption of a purely dipolar magnetic field.
Instead, we considered combinations of dipolar and octupolar components.
This combination was chosen due to the fall off of the octupole component strength with distance from the surface,
thereby providing a magnetic field more concentrated close to the NS
than a quadrupole component of similar strength.

We computed a series of models,
characterized by either a sole low dipolar field (at $3 \times 10^{12} \text{G}$)
or a low dipolar field in addition to a stronger octupolar component ($\sim 3, \ 10$ larger).
We first investigated the solutions by assuming that radiation is dominated by the lower opacity X-mode photons,
and we indeed found a super-Eddington solution
with $L \sim 10^{39} \text{erg s}^{-1}$ and $L \sim 10^{40} \text{erg s}^{-1}$ is always possible.

With respect to models based on a pure dipole field,
we find that when an octopolar component is accounted for,
the accretion column height is lower for a given luminosity.
We find that in order to compensate the decrease in height,
the effective temperature of the accretion column sinking region, $T_{\text{eff}}$,
is larger in models with octupolar components.
Typically, $T_{\text{eff}}$ is in the range $\sim 3 - 15 \ \text{keV}$,
with the peak temperatures at the higher end of the sinking region.
This thermal component is not observed directly
as the radiation escaping the sinking region must pass through a region of free-falling material
inside the accretion column.
It may yet then be reprocessed in a thick accretion curtain \citep{Mushtukov2017}.

This accretion column model relies on the \cite{Ghosh1978} disk model,
specifically in order to derive the truncation radius of the disk
as well as the penetration depth of the flow into the magnetosphere.
One can question to what extent the standard disk model parameters of \cite{Ghosh1978}
affect the accretion column properties
as well as more generally the existence of a solution for a given set of model parameters.
We investigated this aspect, and found that an accurate value of the truncation radius is almost irrelevant,
while the assumption regarding the penetration depth is more crucial (see \S \ref{sec:Disk Model}).
In particular,
a variation of 30\% in the depth can decrease the base linear size of the column by an order of magnitude.

Another limitation concerns the way in which we treat the curvature of the field lines
that constitute the boundary of the accretion column.
A more detailed calculation, which is beyond the purpose of this particular paper,
may account for the influence of the geometry changing with $h$
on the basic hydrodynamical equations (see \citealt{Canalle2005}).

The accretion column properties also depend on the assumed
fraction of X-mode photons present in the radiation field. 
We investigated the impact of this,
by introducing a fixed fraction of O-mode photons (see \S \ref{sec:mixed polarization}). 
Although we found variations in the X-mode fraction introduce differences in the accretion column properties
for most of the magnetic field strengths considered here,
this does not lead to a significant variation in the maximum luminosity curve. 

In general, the opacity local to the NS surface is a decisive factor
in allowing for the super-Eddington luminosities observed in ULXs ($L > 10^{39} \text{erg s}^{-1}$).
However, the geometry of the accretion column footprint on the surface of the NS
is also of crucial importance in determining the column properties.
In particular, we found that the thickness of the accretion column has a significant effect
on the luminosity (see \ref{sec:Disk Model}).
Indeed, we found a greater maximum luminosity for a given magnetic field strength
compared with the one calculated by \cite{Mushtukov2015b} 
in part due to our different approach to calculating the accretion column thickness.

Accretion columns with luminosity $L \sim 10^{41} \text{erg s}^{-1}$ are in principle obtainable 
with our modelling.
However, in order to avoid the propeller regime for a source with pulse period of the order of $\sim 1 \text{s}$,
the dipole field strength must be sufficiently low ($\sim 10^{13} \text{G}$).
A low strength dipole component ($B_{\text{dip}} \sim 10^{12} \text{G}$) together with high accretion luminosity
results in a thick accretion disk,
as already noticed by \cite{Israel2017b}.
This is in contradiction with the assumption of a thin accretion disk of our model.
Consequently, the problem of an upper limit to the luminosity related to the strength of the dipole component remains.

We applied our model to the two sources 
NGC 5907 ULX-1 and NGC 7793 P13 (see \S \ref{sec:application}).
The necessity of a multipolar magnetic field configuration is different for each source,
once beaming is taken into account.

For NGC 7793 P13,
the observed luminosity is $L \approx 1.6 \times 10^{40} \text{erg s}^{-1}$.
When taken face value, this luminosity level is too large to be compatible with the calculation of
\cite{Mushtukov2015b}
since it would require a magnetic field so high that the source would be deep in the propeller regime.
On the other hand, according to our model, 
the lowest observed flux levels are compatible with a purely dipolar configuration
of strength $B_{\text{dip}} \approx 7.3 \times 10^{12} \text{G}$, and 
the addition of a stronger octupole component with surface strength 
of $B_{\text{oct}} > 7.3 \times 10^{13} \text{G}$
can explain the whole range of observed luminosities,  up to the peak value of $1.6 \times 10^{40} \text{erg s}^{-1}$.
This latter particular configuration does not conflict with the propeller effect,
nor with the super-Eddington disk accretion. 
Furthermore, it is compatible with the interpretation of the source spin period derivative
according to a simple treatment (see \S \ref{sec:regime of validity}). 
The only problem is that the largest observed luminosities are not compatible 
with the assumption of a geometrically thin disk,
which in principle may be indicative of an overly simplistic disk model.
Other possibilities include a moderate beaming,
in which case the observed flux levels can be reached even for a purely dipolar magnetic field,
with $B_{\text{dip}} \sim 1.4 \times 10^{12} \text{G}$.
However, this configuration requires that the observed spin period derivative
is due to a secular torque,
most of which is produced by the accumulation of material at the disk-magnetosphere interface
during epochs of high flux level.

The PULX NGC 5907 ULX-1 has a much larger peak luminosity of $2.3\times 10^{41} \text{erg s}^{-1}$.
In this case, 
both a super-Eddington disk accretion regime
and the propeller regime can be avoided by invoking a moderate beaming factor of $b \lesssim 0.15$
(figure \ref{fig:NGC5907ULX1}). 
If the source has a dipolar magnetic field $B_{\text{dip}} \approx 3.2 \times 10^{13} \text{G}$
and a slightly larger octupole surface strength $B_{\text{oct}} \gtrsim 9.6 \times 10^{13} \text{G}$,
then the entire observed range of luminosities can be reached.
Even for this source, stronger beaming factors $b \lesssim 0.02$ allow for a reduced accretion luminosity
and therefore make possible to explain the observed flux levels 
with a pure dipole field configuration, 
provided
that, at the same time, it is assumed that most of
the torque that gives rise to the spin period derivative is
accumulated during phases of larger accretion rate.

There is still a number of open issues that needs to be addressed,
before a self consistent explanation of PULXs can be reached. 
As already mentioned,
the presence of multipole magnetic field components can change the properties of the accretion column significantly.
The maximum shock height, $H$, is reduced in comparison to a pure dipole magnetic field case. 
This in turn results in a higher effective temperature, 
which may manifest in the spectral data. 
In principle, a super strong magnetic field would be able to lower the maximum shock height very close to the surface
so that $H \ll R$.
However, since our model assumes the radiation primarily escapes perpendicular to the sinking region,
its validity would become more dubious as $H \rightarrow 0$.
In addition, as the maximum shock height of the column is lowered,
the temperature of the sinking region may exceed $100 \text{keV}$,
whereupon we expect electron-positron pair creation and annihilation to play an increasingly important role
in limiting the temperature of the accretion column
while also increasing the gas pressure (see \citealt{Mushtukov2019}).
A calculation including the gas pressure as well as pair creation and annihilation will be necessary 
for a more accurate description of the accretion column properties.

Several other simplifying assumptions were made in the model presented in this paper.
First, we assumed that the radiation pressure dominates over the gas pressure
in the sinking region of the accretion column. 
Through numerical calculation of several models,
we found this assumption breaks down at the lower layers of the sinking region.
Since we used a power-law ansatz for the velocity profile of the accreting plasma,
the model is not expected to give an accurate picture of the lower layers of the sinking region,
where the plasma flow becomes stagnant and hence the density becomes infinite.
However, the contribution to the luminosity from these lower layers is negligible compared to higher up in the column,
where the radiation pressure does indeed dominate over the gas pressure.

Second, in our calculation of the scattering opacity,
we neglected the contribution from ions in the plasma and the contribution from vacuum polarization effects,
which both become significant exactly in the strong magnetic field regime we consider here
($B \gtrsim 10^{13} \text{G}$).
Additionally, we assumed that a fixed fraction of X-mode photons made up the radiation field throughout the column
and approximated the opacity as an effective scattering opacity (see \S \ref{sec:scattering cross-section}).
A more physically realistic treatment consists of a careful treatment of the scattering between the polarization modes
as well as including mode switching due to resonant scattering.
This will be the focus of our future work in development of the accretion column model.

Finally, we did not take into account the role of energy advection by the accreting plasma
and cooling via neutrino emission.
These processes were studied by \cite{Mushtukov2018}
and are expected to be relevant in the case of very luminous sources
$L \sim 10^{41} \text{erg s}^{-1}$.
In \S \ref{sec:application}, we opted instead to assume the was beamed by some mechanism.
The exact details of admissible beaming factors for each of these sources is beyond the scope of this paper.

\section{Summary}
\label{sec:Conclusions}
We developed a simplified model of the accretion column for strongly magnetized NSs,
building on and altering the model of \cite{Mushtukov2015b}.
Crucially, we relaxed the assumption of a purely dipolar magnetic field,
which we found allows for a larger maximum luminosity.

We found that when a magnetic field configuration with a significantly strong multipolar component is assumed,
the luminosity released in the accretion column is limited only by the accretion rate from the disk.
This, in turn, calls for more refined models of disk accretion and disk-magnetospheric interaction
at the near-Eddington regime.

We applied the model to two PULXs, NGC 5907 ULX-1 and NGC 7793 P13,
and discussed how their observed properties (luminosity and spin period derivative)
can be explained in terms of different configurations,
either with or without multipolar magnetic components. 
Generally speaking, the latter scenario is more favorable in case the emission is assumed to be highly beamed.
Although at this level it may be difficult to differentiate further, we notice that
strong multipole components may manifest in the spectra or polarization signal,
an issue that we plan to investigate further in following work.

\section*{Acknowledgements}

NB acknowledges STFC for support through a PhD fellowship. 
The work of RT is partially supported by the Italian Ministry
for University and Research through grant PRIN 2017LJ39LM.
We thank A. Mushtukov for discussions and suggestions.
We also thank G. L. Israel for kindly reading the manuscript and providing comments. 
We thank an anonymous referee for very constructive comments on the draft.

\section*{Data Availability}

Data available on request.




\bibliographystyle{mnras}
\bibliography{ref}

\begin{thebibliography}{}
\makeatletter
\relax
\def\mn@urlcharsother{\let\do\@makeother \do\$\do\&\do\#\do\^\do\_\do\%\do\~}
\def\mn@doi{\begingroup\mn@urlcharsother \@ifnextchar [ {\mn@doi@}
  {\mn@doi@[]}}
\def\mn@doi@[#1]#2{\def\@tempa{#1}\ifx\@tempa\@empty \href
  {http://dx.doi.org/#2} {doi:#2}\else \href {http://dx.doi.org/#2} {#1}\fi
  \endgroup}
\def\mn@eprint#1#2{\mn@eprint@#1:#2::\@nil}
\def\mn@eprint@arXiv#1{\href {http://arxiv.org/abs/#1} {{\tt arXiv:#1}}}
\def\mn@eprint@dblp#1{\href {http://dblp.uni-trier.de/rec/bibtex/#1.xml}
  {dblp:#1}}
\def\mn@eprint@#1:#2:#3:#4\@nil{\def\@tempa {#1}\def\@tempb {#2}\def\@tempc
  {#3}\ifx \@tempc \@empty \let \@tempc \@tempb \let \@tempb \@tempa \fi \ifx
  \@tempb \@empty \def\@tempb {arXiv}\fi \@ifundefined
  {mn@eprint@\@tempb}{\@tempb:\@tempc}{\expandafter \expandafter \csname
  mn@eprint@\@tempb\endcsname \expandafter{\@tempc}}}

\bibitem[\protect\citeauthoryear{{Bachetti} et~al.,}{{Bachetti}
  et~al.}{2014}]{Bachetti2014}
{Bachetti} M.,  et~al., 2014, \mn@doi [\nat] {10.1038/nature13791}, \href
  {https://ui.adsabs.harvard.edu/abs/2014Natur.514..202B} {514, 202}

\bibitem[\protect\citeauthoryear{{Basko} \& {Sunyaev}}{{Basko} \&
  {Sunyaev}}{1976}]{Basko1976}
{Basko} M.~M.,  {Sunyaev} R.~A.,  1976, \mn@doi [\mnras]
  {10.1093/mnras/175.2.395}, \href
  {https://ui.adsabs.harvard.edu/abs/1976MNRAS.175..395B} {175, 395}

\bibitem[\protect\citeauthoryear{{Becker}}{{Becker}}{1998}]{Becker1998}
{Becker} P.~A.,  1998, \mn@doi [\apj] {10.1086/305568}, \href
  {https://ui.adsabs.harvard.edu/abs/1998ApJ...498..790B} {498, 790}

\bibitem[\protect\citeauthoryear{{Bilous} et~al.,}{{Bilous}
  et~al.}{2019}]{Bilous2019}
{Bilous} A.~V.,  et~al., 2019, \mn@doi [\apjl] {10.3847/2041-8213/ab53e7},
  \href {https://ui.adsabs.harvard.edu/abs/2019ApJ...887L..23B} {887, L23}

\bibitem[\protect\citeauthoryear{{Borghese}, {Rea}, {Coti Zelati}, {Tiengo}  \&
  {Turolla}}{{Borghese} et~al.}{2015}]{Borghese2015}
{Borghese} A.,  {Rea} N.,  {Coti Zelati} F.,  {Tiengo} A.,   {Turolla} R.,
  2015, \mn@doi [\apjl] {10.1088/2041-8205/807/1/L20}, \href
  {https://ui.adsabs.harvard.edu/abs/2015ApJ...807L..20B} {807, L20}

\bibitem[\protect\citeauthoryear{{Borghese}, {Rea}, {Coti Zelati}, {Tiengo},
  {Turolla}  \& {Zane}}{{Borghese} et~al.}{2017}]{Borghese2017}
{Borghese} A.,  {Rea} N.,  {Coti Zelati} F.,  {Tiengo} A.,  {Turolla} R.,
  {Zane} S.,  2017, \mn@doi [\mnras] {10.1093/mnras/stx632}, \href
  {https://ui.adsabs.harvard.edu/abs/2017MNRAS.468.2975B} {468, 2975}

\bibitem[\protect\citeauthoryear{{Canalle}, {Saxton}, {Wu}, {Cropper}  \&
  {Ramsay}}{{Canalle} et~al.}{2005}]{Canalle2005}
{Canalle} J.~B.~G.,  {Saxton} C.~J.,  {Wu} K.,  {Cropper} M.,   {Ramsay} G.,
  2005, \mn@doi [\aap] {10.1051/0004-6361:20052706}, \href
  {https://ui.adsabs.harvard.edu/abs/2005A&A...440..185C} {440, 185}

\bibitem[\protect\citeauthoryear{{Carpano}, {Haberl}, {Maitra}  \&
  {Vasilopoulos}}{{Carpano} et~al.}{2018}]{Carpano2018}
{Carpano} S.,  {Haberl} F.,  {Maitra} C.,   {Vasilopoulos} G.,  2018, \mn@doi
  [\mnras] {10.1093/mnrasl/sly030}, \href
  {https://ui.adsabs.harvard.edu/abs/2018MNRAS.476L..45C} {476, L45}

\bibitem[\protect\citeauthoryear{{Colbert} \& {Mushotzky}}{{Colbert} \&
  {Mushotzky}}{1999}]{Colbert1999}
{Colbert} E. J.~M.,  {Mushotzky} R.~F.,  1999, \mn@doi [\apj] {10.1086/307356},
  \href {https://ui.adsabs.harvard.edu/abs/1999ApJ...519...89C} {519, 89}

\bibitem[\protect\citeauthoryear{{Dall'Osso}, {Perna}, {Papitto}, {Bozzo}  \&
  {Stella}}{{Dall'Osso} et~al.}{2016}]{Dall'Osso2016}
{Dall'Osso} S.,  {Perna} R.,  {Papitto} A.,  {Bozzo} E.,   {Stella} L.,  2016,
  \mn@doi [\mnras] {10.1093/mnras/stw110}, \href
  {https://ui.adsabs.harvard.edu/abs/2016MNRAS.457.3076D} {457, 3076}

\bibitem[\protect\citeauthoryear{{F{\"u}rst} et~al.,}{{F{\"u}rst}
  et~al.}{2016}]{Furst2016}
{F{\"u}rst} F.,  et~al., 2016, \mn@doi [\apjl] {10.3847/2041-8205/831/2/L14},
  \href {https://ui.adsabs.harvard.edu/abs/2016ApJ...831L..14F} {831, L14}

\bibitem[\protect\citeauthoryear{{Ghosh} \& {Lamb}}{{Ghosh} \&
  {Lamb}}{1978}]{Ghosh1978}
{Ghosh} P.,  {Lamb} F.~K.,  1978, \mn@doi [\apjl] {10.1086/182734}, \href
  {https://ui.adsabs.harvard.edu/abs/1978ApJ...223L..83G} {223, L83}

\bibitem[\protect\citeauthoryear{{Harding} \& {Lai}}{{Harding} \&
  {Lai}}{2006}]{Harding2006}
{Harding} A.~K.,  {Lai} D.,  2006, \mn@doi [Reports on Progress in Physics]
  {10.1088/0034-4885/69/9/R03}, \href
  {https://ui.adsabs.harvard.edu/abs/2006RPPh...69.2631H} {69, 2631}

\bibitem[\protect\citeauthoryear{{Illarionov} \& {Sunyaev}}{{Illarionov} \&
  {Sunyaev}}{1975}]{Illarionov1975}
{Illarionov} A.~F.,  {Sunyaev} R.~A.,  1975, \aap, \href
  {https://ui.adsabs.harvard.edu/abs/1975A&A....39..185I} {39, 185}

\bibitem[\protect\citeauthoryear{{Israel} et~al.,}{{Israel}
  et~al.}{2017a}]{Israel2017b}
{Israel} G.~L.,  et~al., 2017a, \mn@doi [Science] {10.1126/science.aai8635},
  \href {https://ui.adsabs.harvard.edu/abs/2017Sci...355..817I} {355, 817}

\bibitem[\protect\citeauthoryear{{Israel} et~al.,}{{Israel}
  et~al.}{2017b}]{Israel2017a}
{Israel} G.~L.,  et~al., 2017b, \mn@doi [\mnras] {10.1093/mnrasl/slw218}, \href
  {https://ui.adsabs.harvard.edu/abs/2017MNRAS.466L..48I} {466, L48}

\bibitem[\protect\citeauthoryear{{Kaaret}, {Feng}  \& {Roberts}}{{Kaaret}
  et~al.}{2017}]{Kaaret2017}
{Kaaret} P.,  {Feng} H.,   {Roberts} T.~P.,  2017, \mn@doi [\araa]
  {10.1146/annurev-astro-091916-055259}, \href
  {https://ui.adsabs.harvard.edu/abs/2017ARA&A..55..303K} {55, 303}

\bibitem[\protect\citeauthoryear{{Kaminker}, {Pavlov}  \&
  {Shibanov}}{{Kaminker} et~al.}{1982}]{Kaminker1982}
{Kaminker} A.~D.,  {Pavlov} G.~G.,   {Shibanov} I.~A.,  1982, \mn@doi [\apss]
  {10.1007/BF00683336}, \href
  {https://ui.adsabs.harvard.edu/abs/1982Ap&SS..86..249K} {86, 249}

\bibitem[\protect\citeauthoryear{{King} \& {Lasota}}{{King} \&
  {Lasota}}{2016}]{King2016}
{King} A.,  {Lasota} J.-P.,  2016, \mn@doi [\mnras] {10.1093/mnrasl/slw011},
  \href {https://ui.adsabs.harvard.edu/abs/2016MNRAS.458L..10K} {458, L10}

\bibitem[\protect\citeauthoryear{{King}, {Davies}, {Ward}, {Fabbiano}  \&
  {Elvis}}{{King} et~al.}{2001}]{King2001}
{King} A.~R.,  {Davies} M.~B.,  {Ward} M.~J.,  {Fabbiano} G.,   {Elvis} M.,
  2001, \mn@doi [\apjl] {10.1086/320343}, \href
  {https://ui.adsabs.harvard.edu/abs/2001ApJ...552L.109K} {552, L109}

\bibitem[\protect\citeauthoryear{{Li} \& {Wang}}{{Li} \& {Wang}}{1999}]{Li1999}
{Li} X.~D.,  {Wang} Z.~R.,  1999, \mn@doi [\apj] {10.1086/306866}, \href
  {https://ui.adsabs.harvard.edu/abs/1999ApJ...513..845L} {513, 845}

\bibitem[\protect\citeauthoryear{{Lyubarskii} \& {Syunyaev}}{{Lyubarskii} \&
  {Syunyaev}}{1988}]{Lyubarskii1988}
{Lyubarskii} Y.~E.,  {Syunyaev} R.~A.,  1988, Soviet Astronomy Letters, \href
  {https://ui.adsabs.harvard.edu/abs/1988SvAL...14..390L} {14, 390}

\bibitem[\protect\citeauthoryear{{Meszaros}}{{Meszaros}}{1992}]{Meszaros1992}
{Meszaros} P.,  1992, {High-energy radiation from magnetized neutron stars}

\bibitem[\protect\citeauthoryear{Mushtukov, Suleimanov, Tsygankov  \&
  Poutanen}{Mushtukov et~al.}{2015}]{Mushtukov2015b}
Mushtukov A.~A.,  Suleimanov V.~F.,  Tsygankov S.~S.,   Poutanen J.,  2015,
  \mn@doi [Monthly Notices of the Royal Astronomical Society]
  {10.1093/mnras/stv2087}, 454, 2539

\bibitem[\protect\citeauthoryear{{Mushtukov}, {Suleimanov}, {Tsygankov}  \&
  {Ingram}}{{Mushtukov} et~al.}{2017}]{Mushtukov2017}
{Mushtukov} A.~A.,  {Suleimanov} V.~F.,  {Tsygankov} S.~S.,   {Ingram} A.,
  2017, \mn@doi [\mnras] {10.1093/mnras/stx141}, \href
  {https://ui.adsabs.harvard.edu/abs/2017MNRAS.467.1202M} {467, 1202}

\bibitem[\protect\citeauthoryear{{Mushtukov}, {Tsygankov}, {Suleimanov}  \&
  {Poutanen}}{{Mushtukov} et~al.}{2018}]{Mushtukov2018}
{Mushtukov} A.~A.,  {Tsygankov} S.~S.,  {Suleimanov} V.~F.,   {Poutanen} J.,
  2018, \mn@doi [\mnras] {10.1093/mnras/sty379}, \href
  {https://ui.adsabs.harvard.edu/abs/2018MNRAS.476.2867M} {476, 2867}

\bibitem[\protect\citeauthoryear{{Mushtukov}, {Ognev}  \&
  {Nagirner}}{{Mushtukov} et~al.}{2019}]{Mushtukov2019}
{Mushtukov} A.~A.,  {Ognev} I.~S.,   {Nagirner} D.~I.,  2019, \mn@doi [\mnras]
  {10.1093/mnrasl/slz047}, \href
  {https://ui.adsabs.harvard.edu/abs/2019MNRAS.485L.131M} {485, L131}

\bibitem[\protect\citeauthoryear{{Rodr{\'\i}guez Castillo}
  et~al.,}{{Rodr{\'\i}guez Castillo} et~al.}{2019}]{Rodriguez2019}
{Rodr{\'\i}guez Castillo} G.~A.,  et~al., 2019, arXiv e-prints, \href
  {https://ui.adsabs.harvard.edu/abs/2019arXiv190604791R} {p. arXiv:1906.04791}

\bibitem[\protect\citeauthoryear{{Sathyaprakash} et~al.,}{{Sathyaprakash}
  et~al.}{2019}]{Sathyaprakash2019}
{Sathyaprakash} R.,  et~al., 2019, \mn@doi [\mnras] {10.1093/mnrasl/slz086},
  \href {https://ui.adsabs.harvard.edu/abs/2019MNRAS.488L..35S} {488, L35}

\bibitem[\protect\citeauthoryear{{Shakura} \& {Sunyaev}}{{Shakura} \&
  {Sunyaev}}{1973}]{Shakura1973}
{Shakura} N.~I.,  {Sunyaev} R.~A.,  1973, \aap, \href
  {https://ui.adsabs.harvard.edu/abs/1973A&A....24..337S} {500, 33}

\bibitem[\protect\citeauthoryear{{Tiengo} et~al.,}{{Tiengo}
  et~al.}{2013}]{Tiengo2013}
{Tiengo} A.,  et~al., 2013, \mn@doi [\nat] {10.1038/nature12386}, \href
  {https://ui.adsabs.harvard.edu/abs/2013Natur.500..312T} {500, 312}

\bibitem[\protect\citeauthoryear{{Tong} \& {Wang}}{{Tong} \&
  {Wang}}{2019}]{Tong2019}
{Tong} H.,  {Wang} W.,  2019, \mn@doi [\mnras] {10.1093/mnras/sty2989}, \href
  {https://ui.adsabs.harvard.edu/abs/2019MNRAS.482.4956T} {482, 4956}

\bibitem[\protect\citeauthoryear{{Turolla}, {Zane}  \& {Watts}}{{Turolla}
  et~al.}{2015}]{Turolla2015}
{Turolla} R.,  {Zane} S.,   {Watts} A.~L.,  2015, \mn@doi [Reports on Progress
  in Physics] {10.1088/0034-4885/78/11/116901}, \href
  {https://ui.adsabs.harvard.edu/abs/2015RPPh...78k6901T} {78, 116901}

\bibitem[\protect\citeauthoryear{{Wang}}{{Wang}}{1996}]{Wang1996}
{Wang} Y.~M.,  1996, \mn@doi [\apjl] {10.1086/310150}, \href
  {https://ui.adsabs.harvard.edu/abs/1996ApJ...465L.111W} {465, L111}

\bibitem[\protect\citeauthoryear{{Zane}, {Turolla}  \& {Treves}}{{Zane}
  et~al.}{2000}]{Zane2000}
{Zane} S.,  {Turolla} R.,   {Treves} A.,  2000, \mn@doi [\apj]
  {10.1086/309027}, \href
  {https://ui.adsabs.harvard.edu/abs/2000ApJ...537..387Z} {537, 387}

\makeatother
\end{thebibliography}




\bsp	
\label{lastpage}
\end{document}